\documentclass{aa}  
\usepackage{graphicx}  
\usepackage{txfonts}
\usepackage{lscape}
\usepackage{amsmath}
\usepackage{epstopdf} 
\usepackage{natbib}
\usepackage{epsfig}
\bibpunct{(}{)}{;}{a}{}{,} 
\newcommand{\Exp}[1]{\text{\large e}^{\,\,#1}}
\newcommand{\eq}[1]{
\begin{equation}
#1
\end{equation}
}

\begin{document}

\title{Wavelets: a powerful tool for studying rotation, activity, and pulsation in \textit{Kepler} and CoRoT stellar light curves}

\author{J. P. Bravo
\and S. Roque
\and R. Estrela
\and I. C. Le\~ao
\and J. R. De Medeiros}

\institute{Universidade Federal do Rio Grande do Norte, Departamento de F\'{\i}sica, 59072-970
Natal, RN, Brazil \\ \email{jbravo@dfte.ufrn.br}
}

\date{Received date / Accepted date}

\authorrunning{Bravo et al.}
\titlerunning{wavelet procedure}

 
 \abstract
   {}
   {The wavelet transform has been used as a powerful tool for treating several problems in astrophysics. In this work, we show that the time--frequency analysis of stellar light curves using the wavelet transform is a practical tool for identifying rotation, magnetic activity, and pulsation signatures. We present the wavelet spectral composition and multiscale variations of the time series for four classes of stars: targets dominated by magnetic activity, stars with transiting planets, those with binary transits, and pulsating stars.}
   {We applied the Morlet wavelet (6th order), which offers high time and frequency resolution. By applying the wavelet transform to the signal, we obtain the wavelet local and global power spectra. The first is interpreted as energy distribution of the signal in time--frequency space, and the second is obtained by time integration of the local map.}
   {Since the wavelet transform is a useful mathematical tool for nonstationary signals, this technique applied to {\it Kepler} and CoRoT light curves allows us to clearly identify particular signatures for different phenomena. In particular, patterns were identified for the temporal evolution of the rotation period and other periodicity due to active regions affecting these light curves. In addition, a beat-pattern signature in the local wavelet map of pulsating stars over the entire time span was also detected.}
   {}



\maketitle


\section{Introduction}
\label{intro}

The CoRoT \citep{2009IAUS..253...71B} and {\it Kepler} \citep{Borucki19022010} space missions produced unique sets of light curves for about 300,~000 stars, with excellent time sampling and unprecedented photometric precision. These data, in addition to the major scientific goals of the missions (asteroseismology and the search for exoplanets) open new perspectives for studying different stellar properties, including rotation, magnetic activity, and binarity. For extracting information from raw signals, several mathematical transformations can be applied, such as Laplace transform \citep{1945-Widder}, Z-transform \citep{Jury-1964}, Wigner distributions \citep{1988-Boashash}, and Fourier transform \citep*{Bochner-1949}, the last being the most widely used. The wavelet transform \citep{1998BAMS...79...61T} is a more recent tool applied for treating a large number of phenomena in different areas, including geophysics, atmospheric turbulence, health (cardiology), and astrophysics. This transformation has a major advantage, since it allows analysis of frequency variations in time of a given signal. Analogous to sunspots and solar photospheric faculae, whose visibility is modulated by stellar rotation, stellar active regions consist of cool spots and bright faculae caused by the magnetic field of the star. Such starspots are well established as tracers of rotation, but their dynamic behavior may also be used to analyze other relevant phenomena, such as magnetic activity and cycles (e.g., \citet{2014A&A...562A.124M,2009A&A...506...41G,1999GeoRL..26.3613W}).

The present work provides a time--frequency analysis using the wavelet transform of different sort of stellar light curves from CoRoT and {\it Kepler} space missions in order to identify particular features associated to rotation, magnetic activity, and pulsation\footnote{The processing of CoRoT and {\it Kepler} light curves is carried out by an I.C.L. routine called \textit{Coroect}, with methods described in \citet{2013A&A...555A..63D}, and wavelet analysis by a J.P.B. routine, with methods described in this paper, both using the Interactive Data Language (IDL).}. This procedure allows us to obtain a distribution of the signal's energy, in time--scale space, from which we can identify the temporal evolution of different phenomena affecting the light curves (such as active regions and possible beats related to pulsations or surface differential rotation). This paper is organized as follows. Section \ref{methodologie} discusses procedures and methods, with an analysis of artificial stationary/nonstationary signals and a description of the wavelet transform. Sect.~\ref{results} presents the primary results, including the main characteristics of the stars studied here, with conclusions in Sect.~\ref{conclusions}.

\section{Procedures and methods}
\label{methodologie}

The light curve obtained from a star can be decomposed into a number of frequencies represented in the power spectrum, which allows us to determine the periodic components of data that may be related to the physical properties of the system. These properties may be, for instance, rotational modulation and several related dynamic phenomena on the stellar surface, pulsation, as well as planetary transits. From the application of Fourier transform to the signal, we can obtain its frequency--amplitude representation. Nevertheless, since the stellar light curves present events not occurring periodically, such as growth and decay of an active region, our interest, in addition to obtaining the spectral composition, which offers an idea of rotation behavior and pulsation modes, is to follow the time--frequency behavior of those events and identify any specific signature to a particular stellar variability even if the light curve presents some kind of noise or singularities. For this, the time localization of the spectral components is necessary and the application of the wavelet transform to the signal will produce its time--frequency representation (hereafter TFR).

\subsection{Fourier transform}\label{transform}

The Fourier transform, named in honor of Joseph Fourier, is the extension of the Fourier series for nonperiodic functions; it decomposes any function into a sum of sinusoidal basis functions. Each of these functions is a complex exponential of a different frequency $\nu$. The Fourier transform is defined as
\eq{
F(\nu)=\int_{-\infty}^{\infty}f(t)\,\Exp{{-2\pi}{i}{\nu}{t}}dt .
}

\begin{figure}[h]
\center
\includegraphics[height=1.0in,width=3.5in]{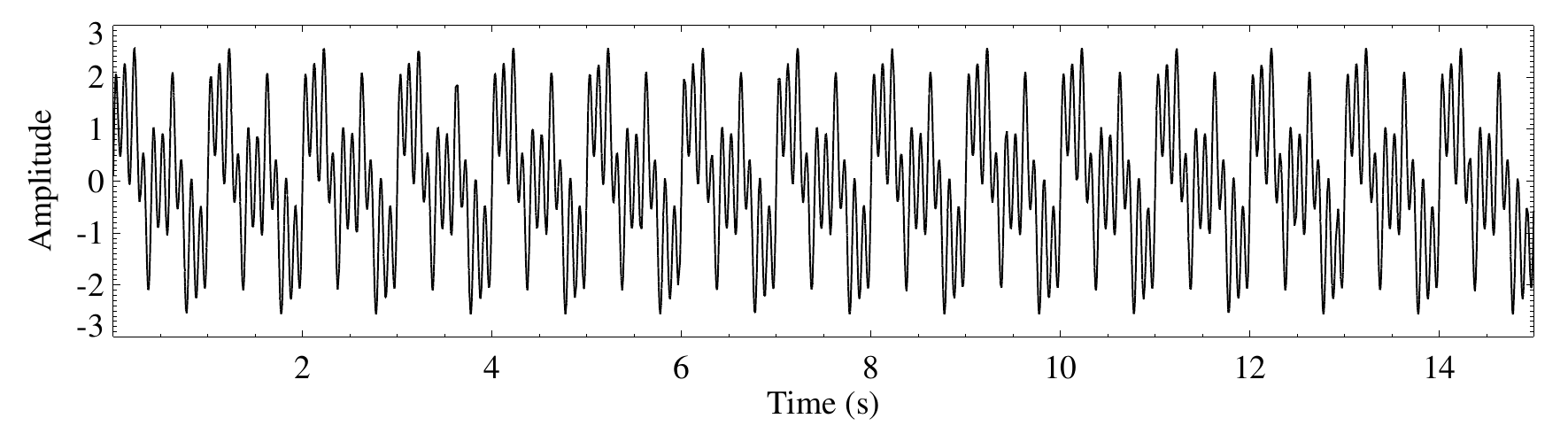}
\includegraphics[scale=0.35]{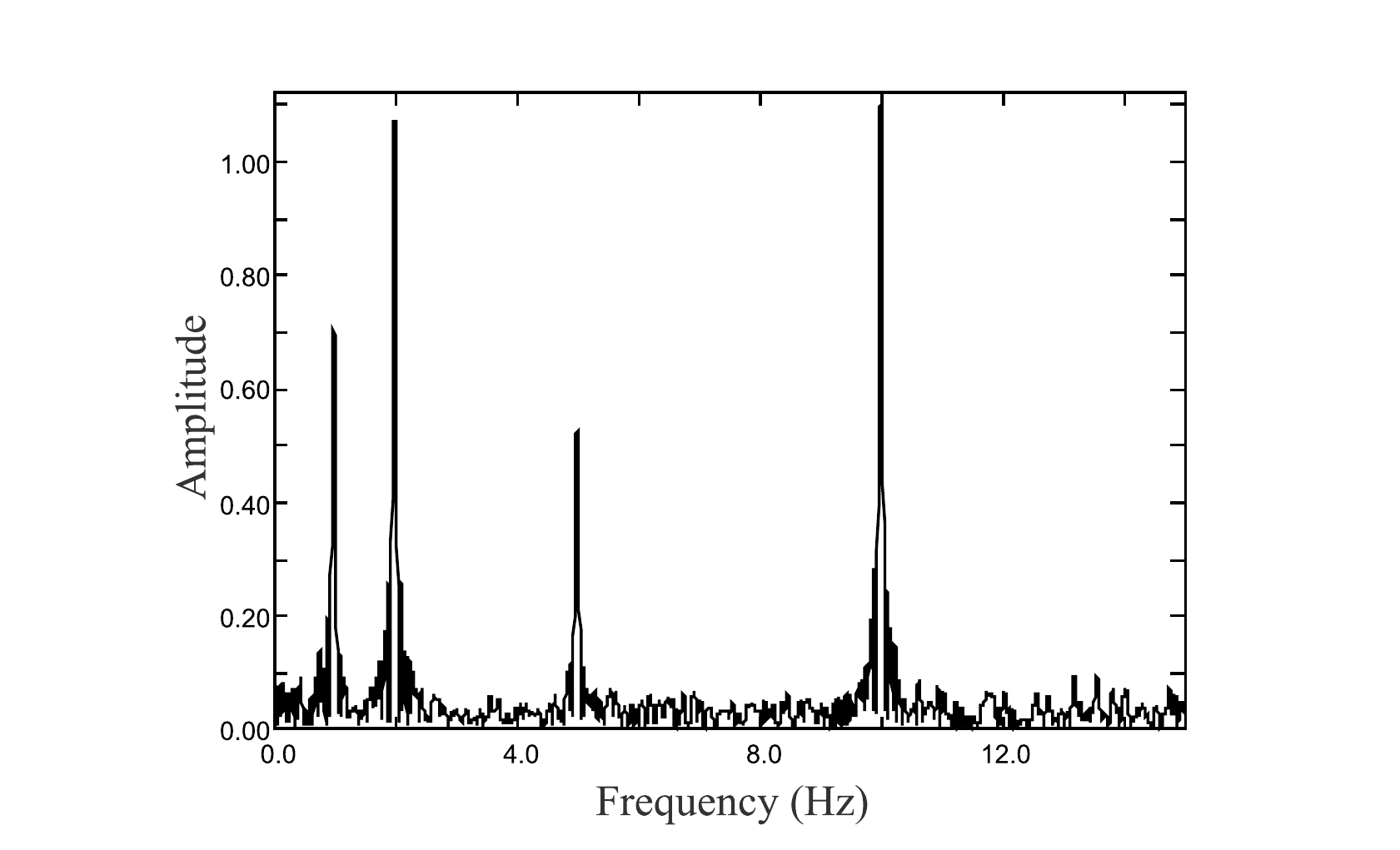}
\caption{Artificial stationary signal composed of sines with different amplitudes and frequencies (1, 10, 5, and 2 Hz) (top panel) and its power spectrum (bottom panel).}
\label{Ssignal}
\end{figure}

The function $F(\nu)$ represents the amount of power inherent in $f(t)$ at frequency $\nu$, providing a frequency-based decomposition of the signal; that is, $F(\nu)$ tells us how much of each frequency exists in the signal, but offers no information on the existence of these frequencies over time. This information is not required when the signal is stationary. As an illustration, we simulated a stationary signal composed of 15-second sines (Fig.~\ref{Ssignal}, top). This signal has different amplitudes and frequencies (1, 10, 5, and 2 Hz) at any given time instant. The power spectrum is obtained by applying the Fourier transform to this signal, as shown in Fig.~\ref{Ssignal} (bottom), where the four spectral components corresponding to frequencies 1, 10, 5, and 2 Hz are identified. 

\begin{figure}
\centering
\includegraphics[height=1.0in,width=3.5in]{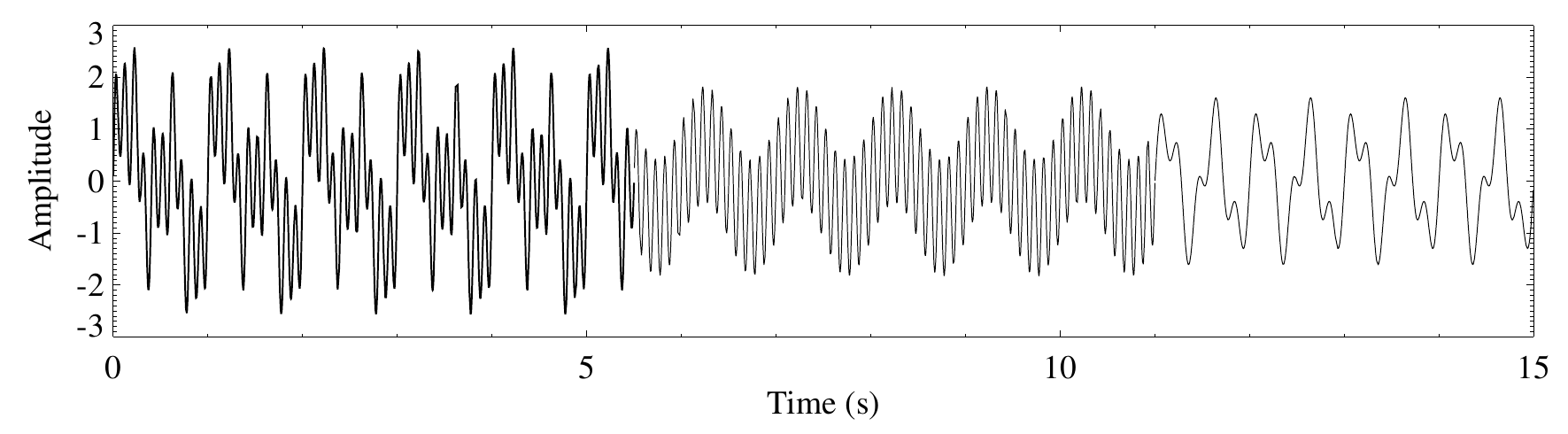}
\includegraphics[scale=0.35]{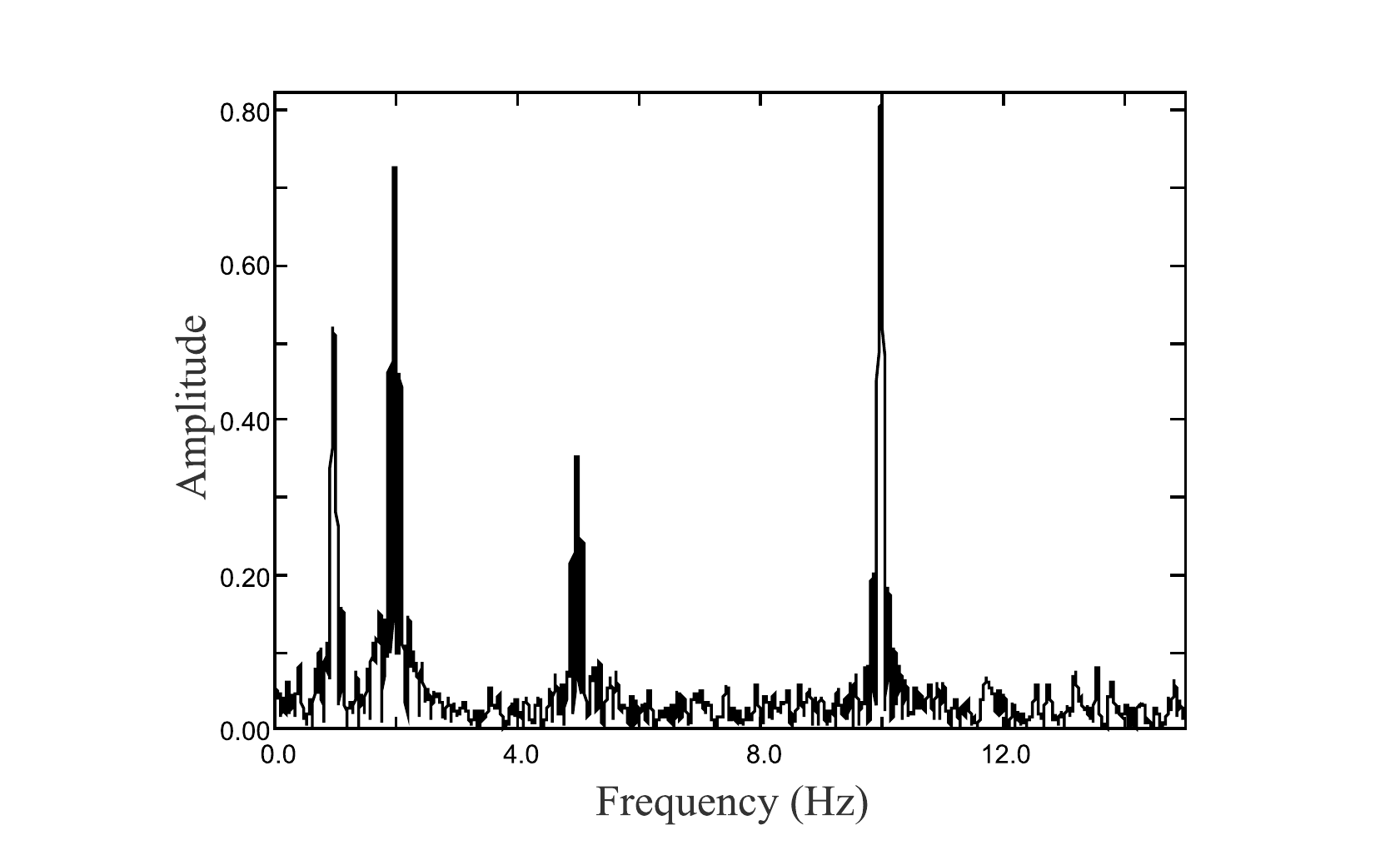}
\caption{Artificial nonstationary signal with four frequencies in three different time intervals (top panel) and its power spectrum (bottom panel).}
\label{NSsignal}
\end{figure}

We also simulated a nonstationary signal with four different frequencies at three different time intervals, shown in Fig.~\ref{NSsignal} (top). In the first interval (up to 5.5 seconds (s)), the signal is composed of the four frequencies; in the second (from 5.5 s to 11 s), only two of the four frequencies are added (1 and 10 Hz); and in the last, the other two frequencies (2 and 5 Hz) make up the final part of the nonstationary signal. The corresponding power spectrum is shown in Fig.~\ref{NSsignal} (bottom).

The two spectra in Figs.~\ref{Ssignal} and \ref{NSsignal} are similar, in the sense that both show four spectral components at exactly the same frequencies (1, 10, 5, and 2 Hz). Nevertheless, although the first simulated signal contains these frequencies at all times, they are present in the second at different time intervals. This is one disadvantage of the Fourier transform, because it provides no information regarding the variation in these frequency components over time. However, globally it provides a more resolved power spectrum than the wavelet transform (Sect.~\ref{wvt}), because it is useful for refining the period determination. On the other hand, the wavelet method allows a better interpretation of physical features (as shown below) prior to considering them for a period refinement.

\citet{Gabor:JIEEE-93-429} modified the Fourier transform, creating the so-called short-term Fourier transform (STFT) or Gabor transform. The mechanism consists of dividing the signal into small enough fixed-length segments, which can be assumed to be stationary. The function to be transformed (signal) is multiplied by a window function $w$, commonly a Gaussian function, and the resulting function is then processed with a Fourier transform to derive the TFR.  Although the STFT has contributed significantly to the study of nonstationary signals, there was still a resolution problem to solve because the STFT does not show what frequency components exist at any given time. Indeed, we only know which frequency band exists at any given time interval \citep{Hubbard-1996}, which is a problem related to the width of the window function used. A wide window gives better frequency resolution but poor time resolution, while a narrow window has the opposite trade-off. This is interpreted as a limit on the simultaneous time--frequency resolution one may achieve (Heisenberg uncertainty principle applied to time--frequency information).

\subsection{The wavelet transform}
\label{wvt}

To overcome the resolution problem, the wavelet technique is a useful tool for analyzing nonstationary and nonperiodic signals, displaying characteristics that can vary in both time and frequency (or scale) \citep*{Burrus-1998}. The central idea of the wavelet is based on \textit{multiresolution analysis}, from which the signal is analyzed at different frequencies with different resolutions showing details of the signal that characterize its different contributions. At low resolution, the details generally characterize large structures and, by increasing the resolution, we obtain more detailed information on the signal. 

In the 1980s, Jean Morlet and Alex Grossman worked together on a mathematical function with two major characteristics: having finite energy and subjected to dilation or compression \citep*{1984-Grossmann-Morlet}. From a convolution between the wavelet and the signal, we can determine how much a section of the signal looks like the wavelet providing the TFR, also called wavelet local power spectrum or wavelet map (hereafter WVM). The analysis uses a function prototype, called \textit{mother wavelet} $\psi(t)$ that generates the other window functions. These functions are called \textit{daughter wavelets} $\psi_{a,b}$ and are defined by translation and dilation (scale) of the mother wavelet $\psi(t)$ as
\eq{
\psi_{a,b}(t)=\frac{1}{\sqrt{a}}\,\,\,\,\psi\left(\frac{t-b}{a}\right),
\qquad a, b \in \Re, a \neq 0,
\label{psi}
}
where $a$ and $b$ are the scale and translation parameters, respectively. Scaling either dilates (large scales) or compresses (small scales) the signal. The constant number $\frac{1}{\sqrt{a}}$ is an energy normalization factor so that the transformed signal will have the same energy at every scale ($E_{signal}=\int_{-\infty}^{+\infty}\mid{f(t)}\mid^{2}{dt}$ where $f(t)$ is a continuous-time signal).

There are two types of wavelet transform: the continuous wavelet transform (CWT) and the discrete wavelet transform (DWT) defined by 
\eq{
CWT{_f}(a,b)=\int{f(t)\,\psi_{a,b}\,dt}=\frac{1}{\sqrt{a}}\,\,\,\int{f(t)\,\psi\left(\frac{t-b}{a}\right)dt}
\label{continous}
}
\noindent{and}
\eq{
DWT{_f}(j,k)=a_{0}\,^\frac{-j}{2}\,\,\,\int{f(t)\,\psi\left(a_{0}^{-j}\,t - k\,b_{0}\right)}\,dt,
\label{discrete}
}
\noindent{respectively, where $j$,$k \in Z$, $a=a_{0}\,^{j}$, $b=k\,b_{0}\,a_{0}\,^{j}$, and $a_{0}$>1 and $b_{0}$>1 are fixed \citep{Foster-1996}. 
The difference between Eqs.~\ref{continous} and \ref{discrete} is that the CWT operates on all possible scales and displacements, whereas the DWT uses a specific set of scales (or frequencies) and displacements (fixed values) \citep{1992tlw..conf.....D}}. In the present work, the CWT is used to achieve periods due to different phenomena, e.g., stellar rotation in some {\it Kepler} and CoRoT stellar light curves.
 
The choice of the mother wavelet is imposed by the information that we want to emphasize in the signal. The most common continuous wavelets are the Morlet and the Mexican hat \citep{Morettin-1999}. The Morlet wavelet is a complex harmonic function contained within a Gaussian envelope as defined by

\eq{
\Psi(t)=e^{-a[\nu(t-b)]^2}\,\,\,e^{-i2\pi\nu(t-b)}
\label{wavelet}
}
where $a$ and $b$ are the scale and translation parameters, respectively, and $\nu$ is related to the order of the wavelet. The second-degree exponential decay of the Gaussian function provides excellent spatial resolution, and its Fourier transform is a Gaussian distribution with very good frequency resolution. In this work, we use the sixth-order Morlet wavelet as the mother wavelet because of its good time localization and frequency resolution.

\begin{figure}
\resizebox{\hsize}{!}{\includegraphics{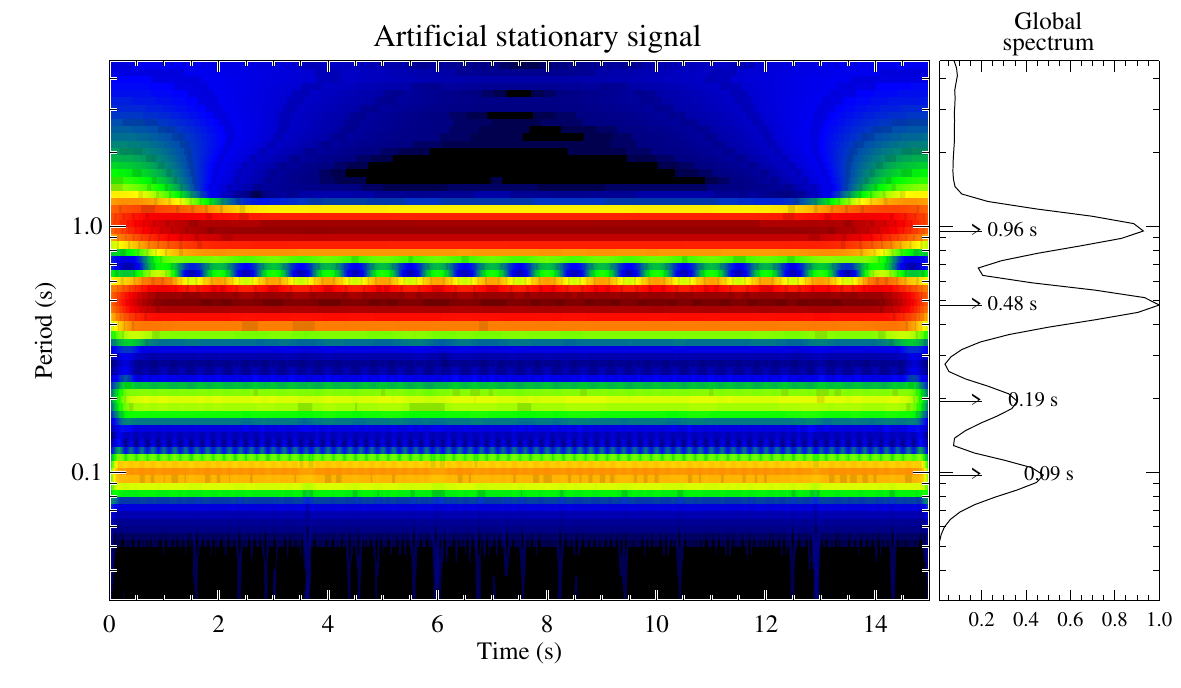}}
\resizebox{\hsize}{!}{\includegraphics{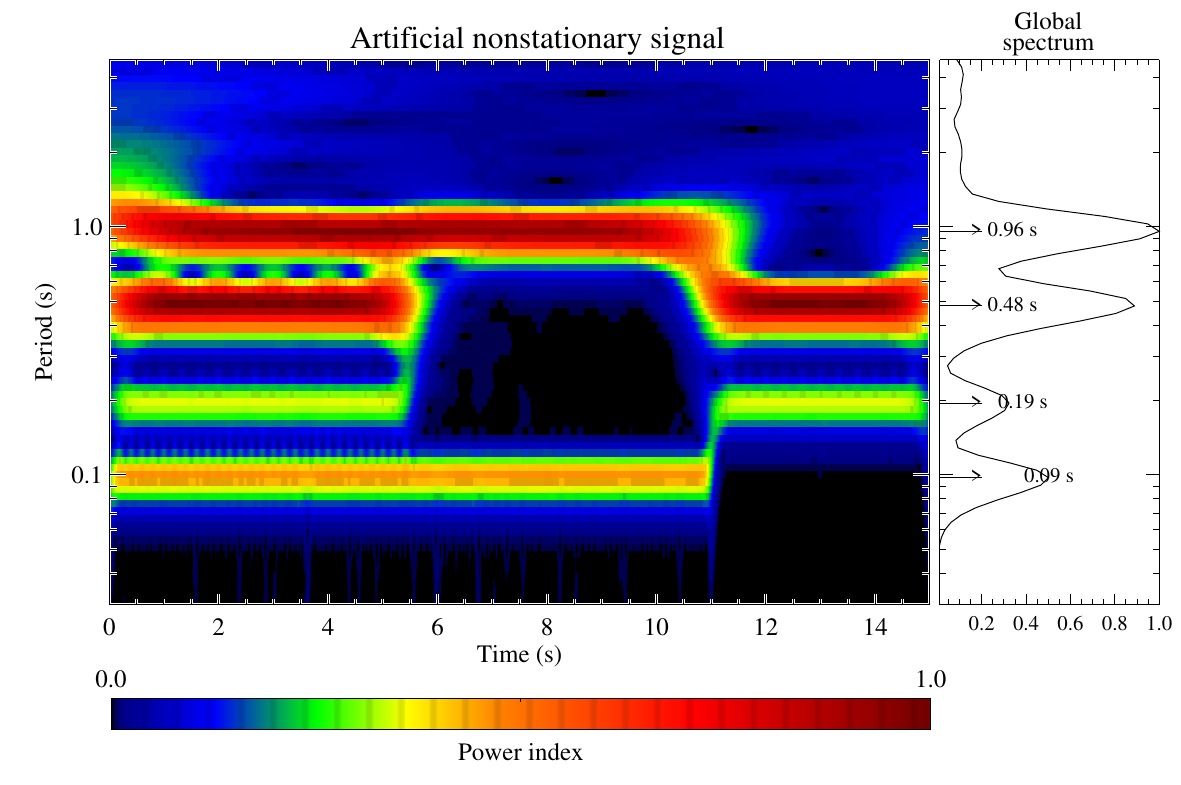}}
\caption{\textit{Top panel}: the wavelet map of the artificial stationary signal of Fig.~\ref{Ssignal}. \textit{Bottom panel}: the wavelet map of the artificial nonstationary signal of Fig.~\ref{NSsignal}. The 6th-order Morlet wavelet was used. Global spectra are illustrated to the right.}
\label{wmp}
\end{figure}

To demonstrate the advantage of the wavelet technique, we applied this method to the artificial stationary and nonstationary signals of Figs.~\ref{Ssignal} and \ref{NSsignal} respectively. Figure~\ref{wmp} shows the resulting WVMs (left panels) and their global wavelet spectra (GWS), which are obtained by time integration of the wavelet local power spectra (right panels). The horizontal and vertical axes correspond to the running time (in seconds) and logarithmic time scale (or the period in seconds (1/$\nu$) where $\nu$ is the frequency in Hz), respectively. At the top, the WVM of the stationary signal shows the presence of four spectral components at all times, which was expected. At the bottom, in the case of the nonstationary signal, we can identify at which time each frequency is present or not. These spectral components are calculated via the GWS and illustrated by an arrow in Fig.~\ref{wmp}. 

Therefore, from the WVM, we can identify which frequencies are predominant in the signal and at which instant they exist or not, resulting in improved accuracy on the timescale. Accordingly, using these WVMs via wavelet transform, we can identify several physical phenomena in stars from their light curves. For example, the technique allows us to determine the rotation period, to identify changes of active regions on the star due to growth or decay of spots and/or to differential rotation, as well as to analyze pulsation.

\section{Results}
\label{results}

In this study, the wavelet method is applied to different {\it Kepler} and CoRoT public stellar light curves, including stars with planetary transit, binary systems, a variable star dominated by magnetic activity, and pulsating stars. Indeed, we present here the results of our analysis for a set of targets listed in different studies of variability reported in the literature, to compare our results with those produced by different procedures. First of all, we analyzed the CoRoT-2, which is a widely studied star, in order to better understand the surface phenomena behavior of this young, spotted yellow dwarf, and also Kepler-4, a Sun-like star presenting low changes in amplitude in its light curve allowing us to consider it as a quiet star. A {\it Kepler} apparently single star (KIC 1995351) dominated by magnetic activity was also analyzed, in the interest of finding similar spot dynamics behavior as in CoRoT-2 with the transit removed\footnote{The planetary and binary transits are removed using the I.C.L. routine based in \citep{2003ApJ...589.1020D} and \citep{2010SPIE.7740E..16T} methods.}. In addition, we applied the wavelet procedure to a {\it Kepler} eclipsing binary system, KIC 7021177, as well as to four pulsating variable stars, two targets observed by CoRoT (CoRoT 105288363 and CoRoT 102918586) and two by {\it Kepler} (KIC 9697825 and KIC 3744571). The light curves of the {\it Kepler} targets observed at a cadence of $\sim$ 30 minutes (with a mean total time span of 1380 days) were reduced with the Pre-Search Data Conditioning (PDC) module of the {\it Kepler} data analysis pipeline, which tries to remove discontinuities, outliers, systematic trends, and other instrumental signatures \citep{2010SPIE.7740E..62T}. Those of the CoRoT stars are provided by the CoRoT public N2 data archive \citep{2006ESASP1306..145B}.
 
\subsection{The Sun}

Because of its proximity, the Sun has become a standard model for studying stars. By analogy with the Sun and the solar magnetic cycles, active regions identified in other stars offer the possibility of studying stellar differential rotation, magnetic activity, dynamic of spots, and cycle variability. In this context, before dealing with our selected sample of stars, we briefly describe the results from the wavelet procedure applied to the total solar irradiance (TSI) time series from 1976 until 2013, including cycles 21-23 and the beginning of cycle 24, obtained from radiometers on different space platforms: HF on Nimbus7, ACRIM I on SMM, ACRIM II on UARS, and VIRGO on SOHO. The composite TSI time series, taken from the World Radiation Center, was expanded back to the minimum in 1976 using a model described in \citet{2004A&ARv..12..273F}.

\begin{figure}
\resizebox{\hsize}{!}{\includegraphics{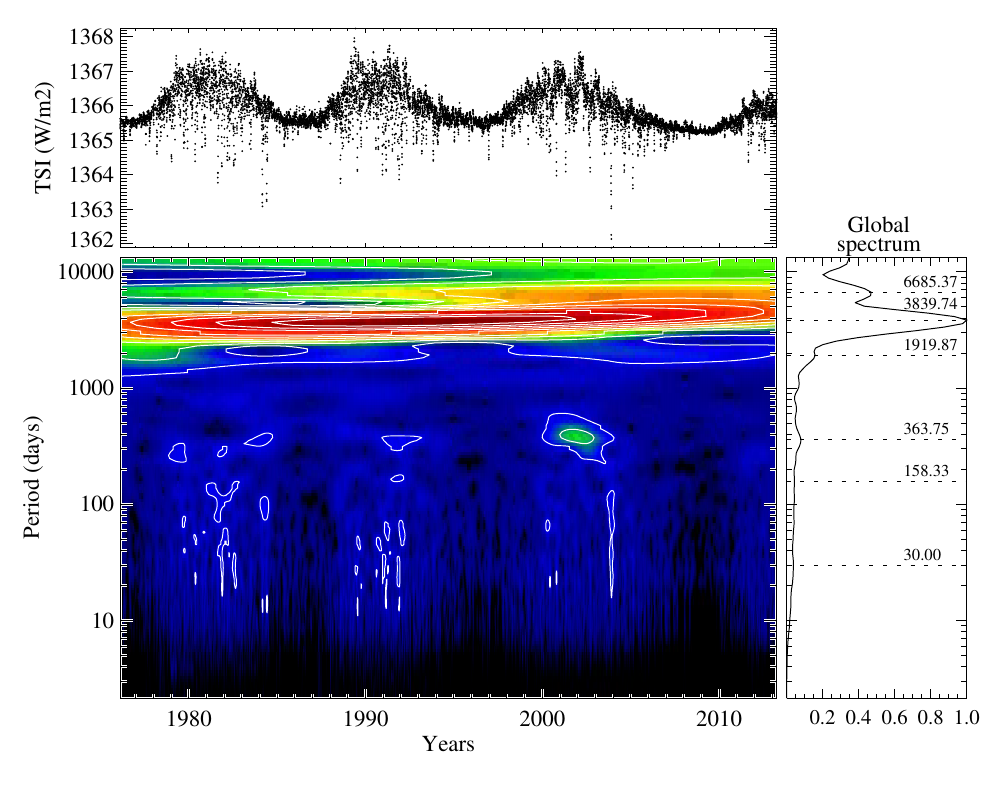}} 
\resizebox{\hsize}{!}{\includegraphics{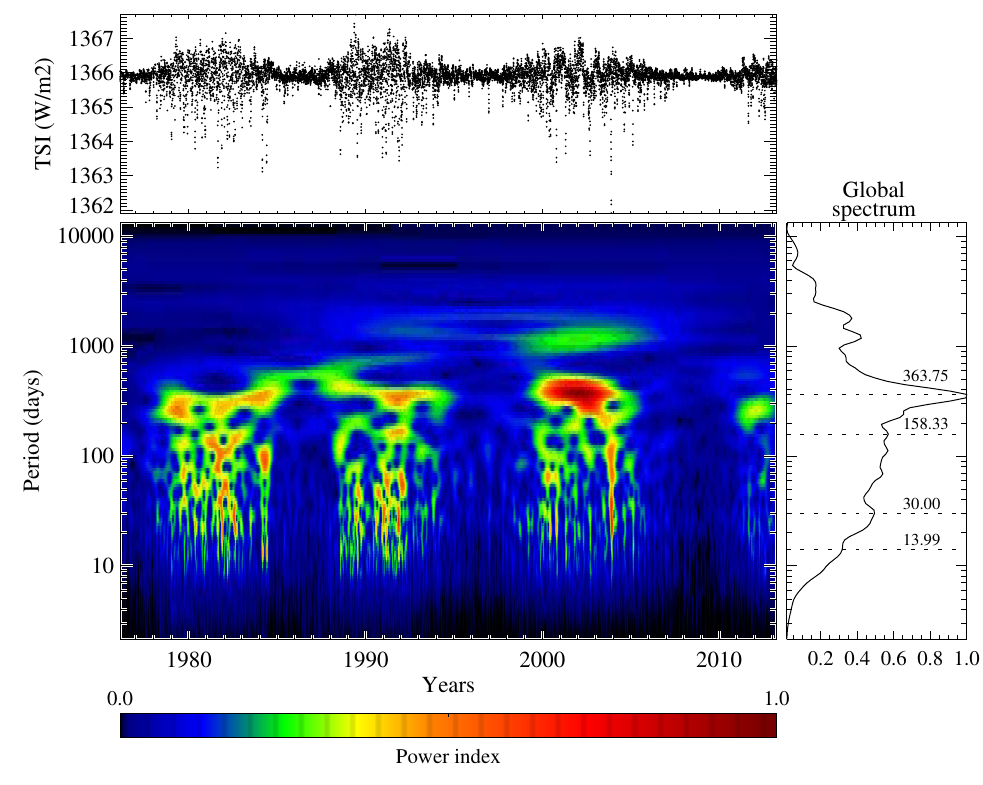}}
\caption{\textit{Top panel}: the composite TSI time series (at the top) and its wavelet local/global power spectra (in the center/to the right) considering long-term contributions. The most dominant feature of the spectrum is the 11-year cycle periodicity ($P = 3840$ days). \textit{Bottom panel}: the long-term contributions are removed. The intermediate- and short-term periodicities are clearly identified: 364 days (annual solar feature related to magnetic fluxes generated deep inside the Sun), 158 days (Rieger-type period), 30 days (solar rotation period) and 14 days (harmonic of the rotation period and associated to solar active regions). Contour levels are 90\%, 80\%, 70\%,..., 20\% and 10\% of the map maximum. The contour levels are not plotted in the bottom panel for better viewing of periods. The 6th-order Morlet wavelet was used.}
\label{wvsun}
\end{figure}

Figure~\ref{wvsun} shows the wavelet analysis for the Sun, where in each panel we present the time series at the top, its local map (modulus of the CWT and normalized to its maximum value) in the center, whose amplitudes are shown in terms of a color contour map, and the GWS (as the weighted average by time span) to the right. The WVM of the top panel exposes the 11-year cycle periodicity (or 3840 days using our method), which is the most dominant feature of the spectrum even if in the GWS we see other periodicities and some subharmonics with lower power. Removing the long-term contributions, the intermediate- and short-term solar periodicities, as well as their changes over the entire time span, are clearly identified (bottom panel of Fig.~\ref{wvsun}). 

The dominant feature in the global spectrum is the 364-day periodicity, which is probably related to the 1.3-year periodicity at the base of the solar convection zone, as reported by \citet{Howe31032000}, and also detected in sunspots areas and sunspots number time series studied using wavelet transforms by \citet{2002A&A...394..701K}, leading to an association of this period with an annual solar feature caused by magnetic fluxes generated deep inside the Sun. The other interesting features in the spectrum are the 158-day, 30-day, and 14-day periodicities. The 158-day flare occurrence period, called the Rieger-type period \citep{1984Natur.312..623R}, is stronger in cycle 21 but weaker for the subsequent cycles, and almost absent in the cycle 24. \citet{2002A&A...394..701K} propose that Rieger-period is the third harmonic of the 1.3-year period. We also obtained the solar rotation period of 30 days, which is more evident in cycle 21 because of maximum activity but persists over the next three cycles. The identified 14-day periodicity seems to be a harmonic of the 30-day variation. \citet{1990SoPh..130..369D} demonstrate that such a cycle is also associated to the active regions located opposite each other in solar longitude. The solar periodicities issued from the present analysis are in close agreement with those obtained by different authors, on the basis of different procedures for the treatment of the total solar irradiance (e.g., \citet{1997Sci...277.1963W,1999GeoRL..26.3613W,1998GeoRL..25.4377F}).

\subsection{Stars with transiting planet}

One of the first planets detected with the CoRoT satellite, during its first long run in the galactic center direction (LRc01, time base 142 days), was CoRoT-Exo-2b, a hot Jupiter with a 1.743-day orbit around a main-sequence G7V star. Because its stellar mass, radius, and effective temperature are comparable to those of the Sun and because it is to the ZAMS \citep{2008A&A...482L..25B}, which is possibly younger than 0.5 Gyr, CoRoT-2 (\object{CoRoT 101206560}, 2MASS 19270649+0123013) has become a laboratory for our understanding of the magnetic activity behavior of the young Sun. The physical parameters of the star and the planet's characteristics were determined by \citet{2008A&A...482L..21A} and \citet{2008A&A...482L..25B}. Photometric analysis shows that modulation of the star was related to two active longitudes initially on opposite hemispheres, i.e., separated by $\sim{180}^\circ$. The first does not appreciably migrate, showing a rotation period of 4.522 $\pm$ 0.0024 days, and the other slowly migrates (retrograde migration) with a rotation period of 4.554 d \citep{2009A&A...493..193L}. 

\begin{figure}
\resizebox{\hsize}{!}{\includegraphics{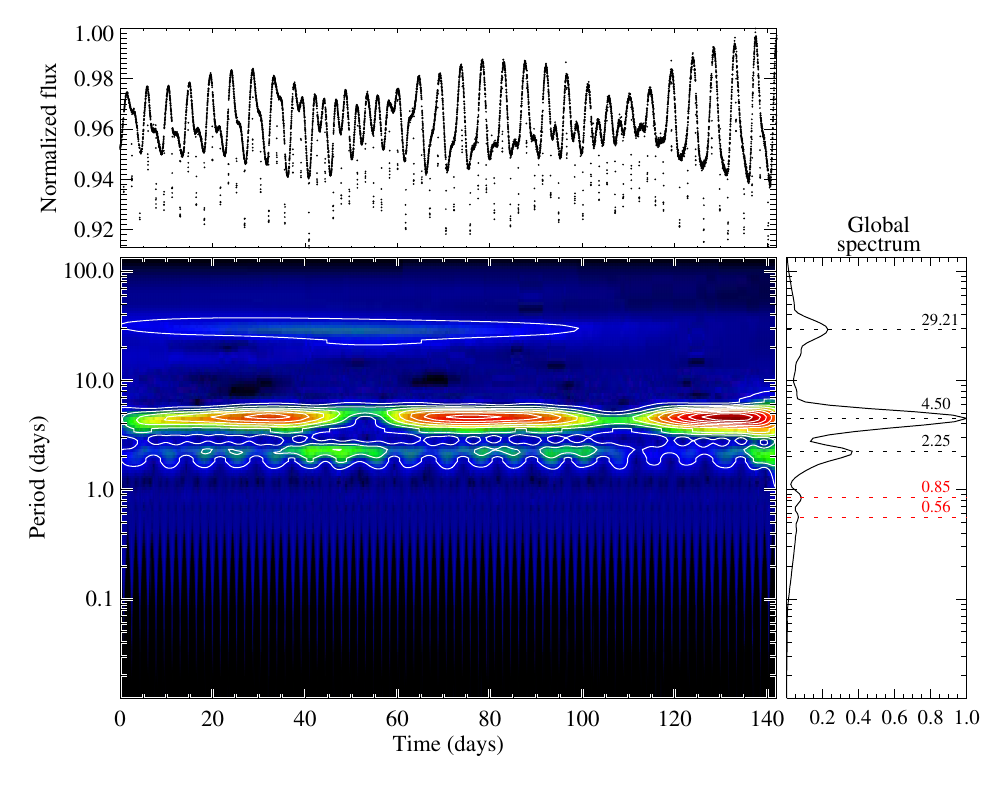}} 
\resizebox{\hsize}{!}{\includegraphics{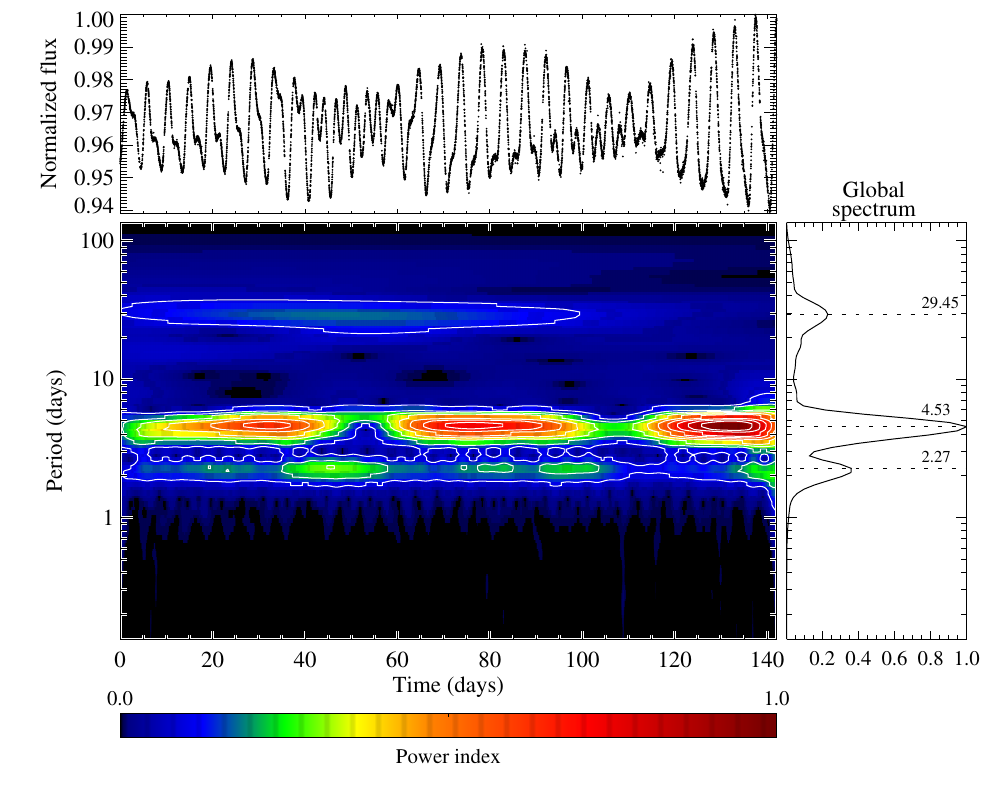}}
\caption{\textit{Top panel}: light curve of CoRoT-2 with transiting planet (at the top), its wavelet map (in the center) and global spectrum (to the right). \textit{Bottom panel}: light curve of CoRoT-2 with transits removed (at the top), its wavelet map (in the center) and global spectrum (to the right). Contour levels are 90\%, 80\%, 70\%,..., 20\% and 10\% of the map maximum. The 6th-order Morlet wavelet was used.}
\label{wmp2}
\end{figure}

Figure~\ref{wmp2} shows the wavelet map with the associated global spectrum for this star. In the WVM of the top panel, the transits have a fine and deep droplet form. The periods related to the transit are indicated in the GWS by a red dashed line (0.85~d and 0.56~d) in the top right hand panel of Fig.~\ref{wmp2}, which no longer appear in the GWS once the transit is removed (bottom right panel). Indeed, these are aliases of the planet orbital period. Because the periodicity associated to the magnetic activity is mixed with the energy contribution of the transits in the WVM, this hides the real orbital period and outstrips their aliases.
The transits are removed because they can alter the periodogram when the orbital and the rotational periods (or their aliases) are synchronized or at least very close, preventing us from visualizing the persistence of the predominant periods. In this case we do not see any significant differences between both WVMs, but in some cases such as the binary system in section \ref{BS} which presents deeper transits, it is necessary to remove them. 

As seen in the WVM of the CoRoT-2 light curve, after removing the planet transit there is a clear signature showing the persistence over time of two semi-regular ``dune'' ranges (assemblages of color levels) representing the two predominant periods, which are calculated by a time integration of the local map and illustrated by a black dashed line in the GWS. Thus, we have 4.53~d as the rotation period and 2.27~d (approximately half of the main period) associated to spot emergence on opposite hemispheres of the star, most likely caused by differential rotation. This feature is therefore considered an indicator of rotational modulation related to the starspots. These periods are in accordance with the results obtained by Lanza's method \citep{2009A&A...493..193L} and also compared using the Lomb-Scargle method, which gives us a main period of 4.528~d and a second of 2.271~d, allowing us to adopt this type of signature in the WVM as a magnetic activity signature.
Also calculated via Lomb-Scargle, there is another period of 29.45~d, which could represent the variation in intensity of the second spot area or to be related to cyclic oscillation of the total spotted area, as reported by \citet{2009A&A...493..193L}.

\begin{figure}
\resizebox{\hsize}{!}{\includegraphics{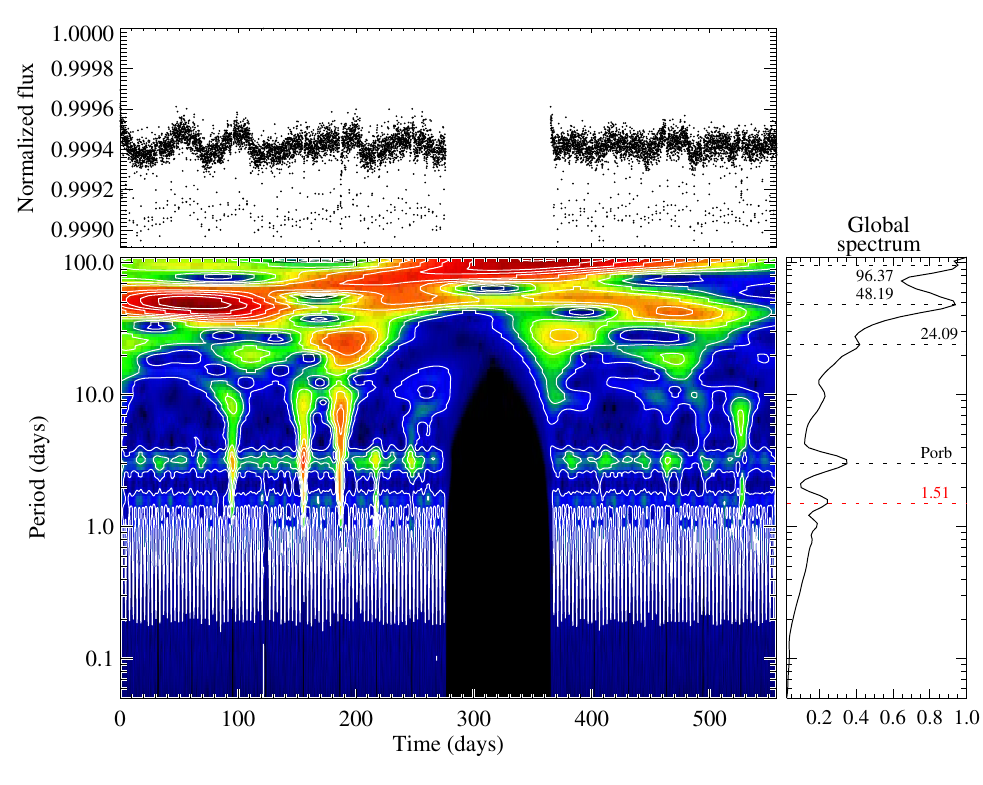}} 
\resizebox{\hsize}{!}{\includegraphics{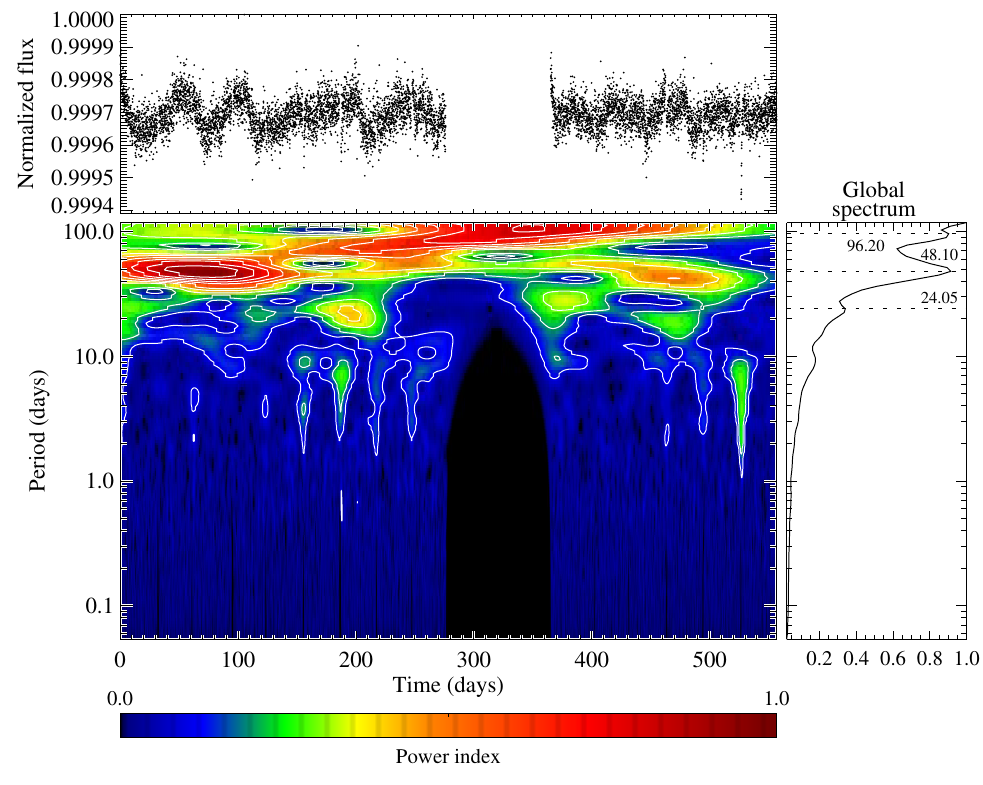}}
\caption{\textit{Top panel}: light curve of Kepler-4 with transiting planet (at the top), its wavelet map (in the center) and global spectrum (to the right). \textit{Bottom panel}: light curve of Kepler-4 with transits removed (at the top), its wavelet map (in the center) and global spectrum (to the right). Contour levels are 90\%, 80\%, 70\%,..., 20\% and 10\% of the map maximum. The 6th-order Morlet wavelet was used.}
\label{wmp4}
\end{figure}

In contrast to the CoRoT-2 star, the second star with planetary transit analyzed in the present study, Kepler-4 (\object{KIC 11853905}, 2MASS 19022767+5008087), a G0-type star, is slightly brighter and is regarded in this work as a quiet star. Its planet Kepler-4b discovered in 2010 has the size of Neptune and orbits its host star in 3.21 days \citep{2011ApJ...736...19B}. Figure~\ref{wmp4} shows the wavelet map with the associated global spectrum for Kepler-4. We used here the {\it Kepler} Quarters 5-7 and Quarters 9-10, which yield 468-day time series once added. After removing the planetary transit (bottom panel), resulting in a cleaner wavelet map, we observe that the orbital period $P_{orb} = 3.01$~d (obtained with our method) and its alias (red dashed line in the GWS of the top panel) no longer appear, thereby making three periodicities evident. Because the amplitude variations are very small, we can hardly be certain which periodicity is related to rotation. A first guess is that rotation is associated with the 48.10-day periodicity, and the others with their harmonics. In this case, no evident magnetic activity signature is identified, and the two dune ranges do not appear in the wavelet map for this quiet star.  

\subsection{Variable star dominated by magnetic activity}
 
\begin{figure}
\resizebox{\hsize}{!}{\includegraphics{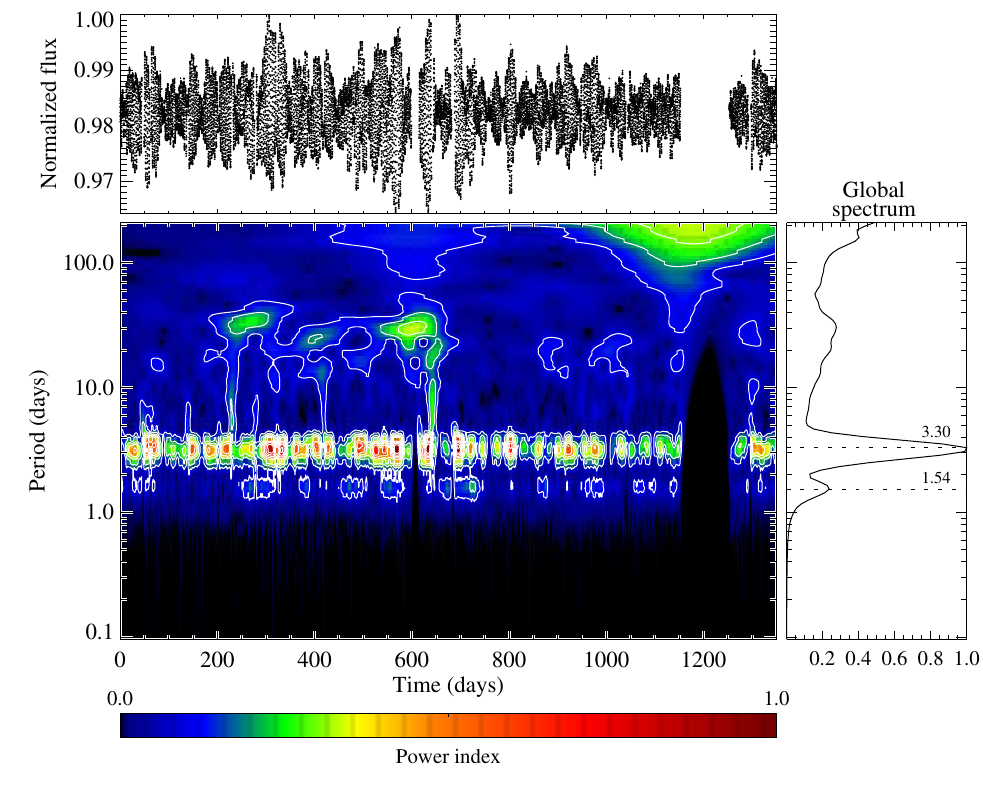}}
\caption{Light curve (at the top) of {\it Kepler} star KIC 1995351, its local map (in the center), and its global wavelet spectrum (to the right). Contour levels are 90\%, 80\%, 70\%,..., 20\% and 10\% of the map maximum. The 6th-order Morlet wavelet was used.}
\label{wmpKIC5351}
\end{figure}

Here we present an example of an apparently single active star, \object{KIC 1995351} ($ra = 19^{h}04^{m}23.2^{s}$, $dec$ = +37$\degr$27$\arcmin$18.0$\arcsec$, J2000) in the search for a magnetic activity signature comparable to that of CoRoT-2. In fact, the light curve of this star shows significant variability features, which, in principle, could be associated to the pulsation or to rotational modulation caused by active regions. Its wavelet map with the associated global spectrum is displayed in Fig.~\ref{wmpKIC5351}, confirming the hypothesis that it is a fast rotator with many active regions, reflected by the semi-regular pattern observed in the light curve. Two main periods persist in the local map over the entire time span, both also evident in the GWS. The most significant, around 3.30~d, is related to the rotation, and the second, almost equal to half of the primary period, that is, 1.54~d, is associated to active regions that could be on opposite sides of the star, be growing and decaying, or be migrating, forming a double dip in the light curve (easily identified by visual inspection in some quarters as a local feature). One important aspect to underline here is that the period 1.54~d can be a potential indicator for two or more active regions contributing to the signal, with a period close to the main value of 3.30~d and suffering relative changes from one to another. From Lomb-Scargle in a prewhitening approach, \citet{2013arXiv1308.1508R} found two main periodicities for this case: $P_{1} = 3.24$~d and $P_{2} = 3.57$~d.  

\subsection{Binary system}
\label{BS}

\begin{figure}
\resizebox{\hsize}{!}{\includegraphics{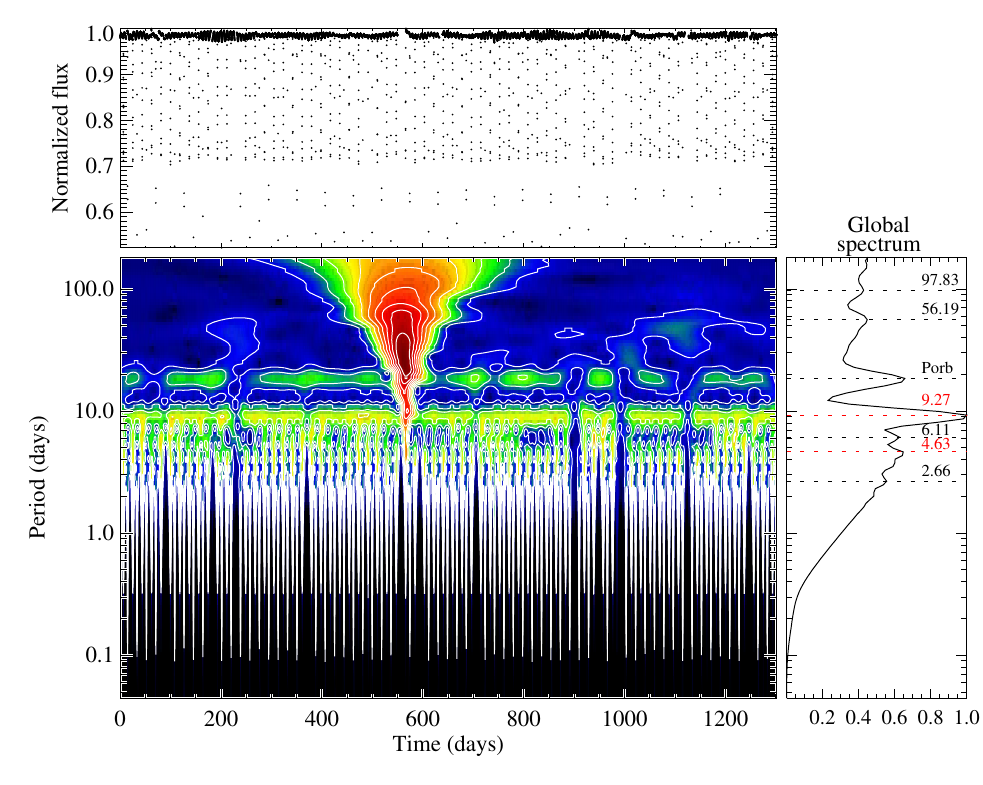}}
\resizebox{\hsize}{!}{\includegraphics{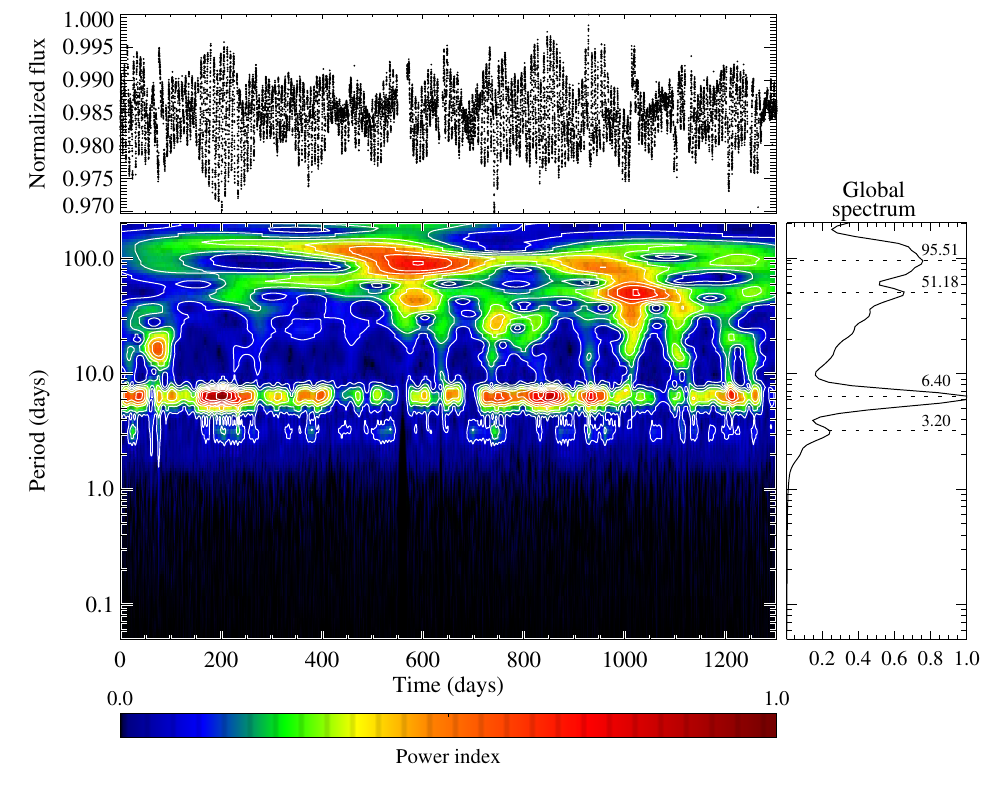}}
\caption{\textit{Top panel}: light curve of KIC 7021177 with binary transits (at the top) and its local/global wavelet map (in the center/to the right). \textit{Bottom panel}: light curve of KIC 7021177 with eclipses removed (at the top) and its local/global wavelet map (in the center/to the right). Contour levels are 90\%, 80\%, 70\%,..., 20\% and 10\% of the map maximum. The 6th-order Morlet wavelet was used.}
\label{wmp1177}
\end{figure}

In this section we present the wavelet analysis for the {\it Kepler} binary system \object{KIC 7021177} (2MASS 19103289+4231509), classified as an eclipsing binary by \citet{2011AJ....141...83P} and studied by \citet{2012BlgAJ..18c..81D}. Its wavelet map is displayed in Fig.~\ref{wmp1177} with the associated global spectrum. The orbital period $P_{orb} = 18.54$~d calculated via wavelet procedure and illustrated in the GWS (top panel) conforms with $P_{orb} = 18.6$~d of \citet{2012BlgAJ..18c..81D}. Also, we observe some possible aliases (9.27~d and 4.63~d) of the transit period in both the WVM and the GWS (red dashed line). To search for stable periods, namely those that are persistent along the entire light curve, we removed the eclipses (bottom panel), whose depths are greater than the amplitudes of the rotational modulation contribution, distorting both the WVM and the GWS. We observe that the aliases are no longer present after the eclipses are removed. The regular changes in the light curve are represented by two semi-regular and continuous dune ranges over the 1300-day window in the WVM, corresponding to the remaining 6.40-day and 3.20-day periodicities. The first is associated with rotational modulation caused by spots, in agreement with the rotation period computed by \citet{2012BlgAJ..18c..81D}, whereas the second is the second harmonic that may be caused by active regions located $180^{\circ}$ apart on the stellar surface. As we have seen previously, this feature was also observed for the Sun and CoRoT-2 star. The light curve shows that some active regions emerge and fade during the entire coverage period with lower amplitude variation, which is characterized in the WVM by the dune ranges and their power index variations. Two other periodicities at 95.51~d and 51.18~d are also present in the WVMs, but it seems that both are caused by the recurrent gaps in the light curve.   

\subsection{Pulsating stars}

Changes in the luminosity of stars are also caused by fluctuations in stellar radius. This phenomenon is present in many intrinsic variable stars, such as RR Lyrae, Cepheids, and Delta Scuti, producing large and rather regular variations in amplitude in their light curves. Some variable stars, such as Gamma Doradus stars ($\gamma$ Dor), are nonradial pulsators and have smaller pulsation amplitudes. Here we analyze two types of pulsating stars, \object{CoRoT 105288363} and KIC 9697825, which are typical examples of RR Lyrae stars, and CoRoT 102918586 and KIC 3744571, presenting the typical behavior of $\gamma$ Dor stars.

\begin{figure}
\resizebox{\hsize}{!}{\includegraphics{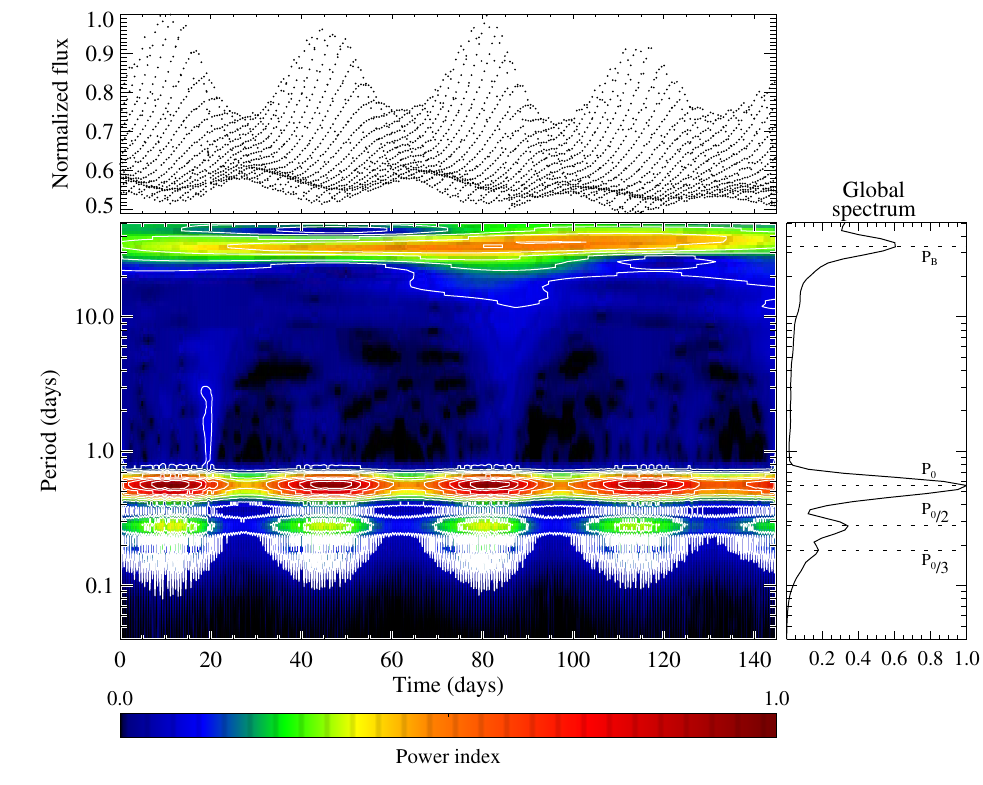}}
\caption{Light curve (at the top) of CoRoT star 105288363, an RRab-type Blazhko RR Lyrae, its local map (in the center) and global wavelet spectrum (to the right). The main periodicities detected are $P_{0} = 0.56$~d, associated to the radial fundamental pulsation period, $\frac{P_{0}}{2} = 0.28$~d and $\frac{P_{0}}{3} = 0.18$~d, the second and third harmonics, and $P_{B} = 33.27$~d, related to the Blazhko period. Contour levels are 90\%, 80\%, 70\%,..., 20\% and 10\% of the map maximum. The 6th-order Morlet wavelet was used.}
\label{wmpRRLYRAE}
\end{figure}

The CoRoT star 105288363 ($ra = 18^{h}39^{m}30.8^{s}$, $dec$ = +7$\degr$26$\arcmin$55.3$\arcsec$, J2000), observed during the second long run in the galactic center direction (LRc02, time base 145 days), is a new RRab-type Blazhko RR Lyrae star (pulsation in the radial fundamental mode), analyzed by \citet{2011MNRAS.415.1577G}. Their results are considered here as a comparison with our findings obtained via the wavelet procedure. The light curve and wavelet map with the associated global spectrum for the referred star are shown in Fig.~\ref{wmpRRLYRAE}. Its local map shows the long-term behavior of pulsation and its stability on low scales (high frequencies) of less than one day, represented by a track associated to the stronger power index. Some harmonics are illustrated by weaker power tracks. These periodicities are indicated in the global spectrum. Period $P_{0} = 0.56$~d (or frequency $f_0 = 1.785$~d$^{-1}$) corresponds to the radial fundamental pulsation period, and the second and third harmonics are $\frac{P_{0}}{2} = 0.28$~d (or $2f_0 = 3.571$~d$^{-1}$) and $\frac{P_{0}}{3} = 0.18$~d (or $3f_0 = 5.556$~d$^{-1}$). Finally, $P_{B} = 33.27$~d (or $f_B = 0.03$~d$^{-1}$) is associated to the Blazhko modulation. We underline that the Blazhko effect is a variation in period and amplitude in RR Lyrae type variable stars (e.g., \citet{2014IAUS..301..241S}). The CoRoT star 105288363 exhibits clearly strong cycle-to-cycle changes in Blazhko modulation. In a continuous time span, 255 pulsations and more than 4 full Blazhko cycles were observed and investigated by \citet{2011MNRAS.415.1577G}. These cycles are clearly observable in the local map displayed in Fig.~\ref{wmpRRLYRAE} in color intensity and shape, forming a beat pattern with some circular and regular dune ranges. These periodicities are in accordance with those calculated by \citet{2011MNRAS.415.1577G} using Fourier analysis ($f_{0} = 1.7623$~d$^{-1}$, $f_{1} = 2.984$~d$^{-1}$, $f_{B} = 0.028$~d$^{-1}$, with $f_{B}$ the Blazhko frequency). Nevertheless when using our method, we do not find the additional period $f_{1}$, considered as an independent mode by \citet{2011MNRAS.415.1577G}, possibly owing to its very low amplitude.

\begin{figure}
\resizebox{\hsize}{!}{\includegraphics{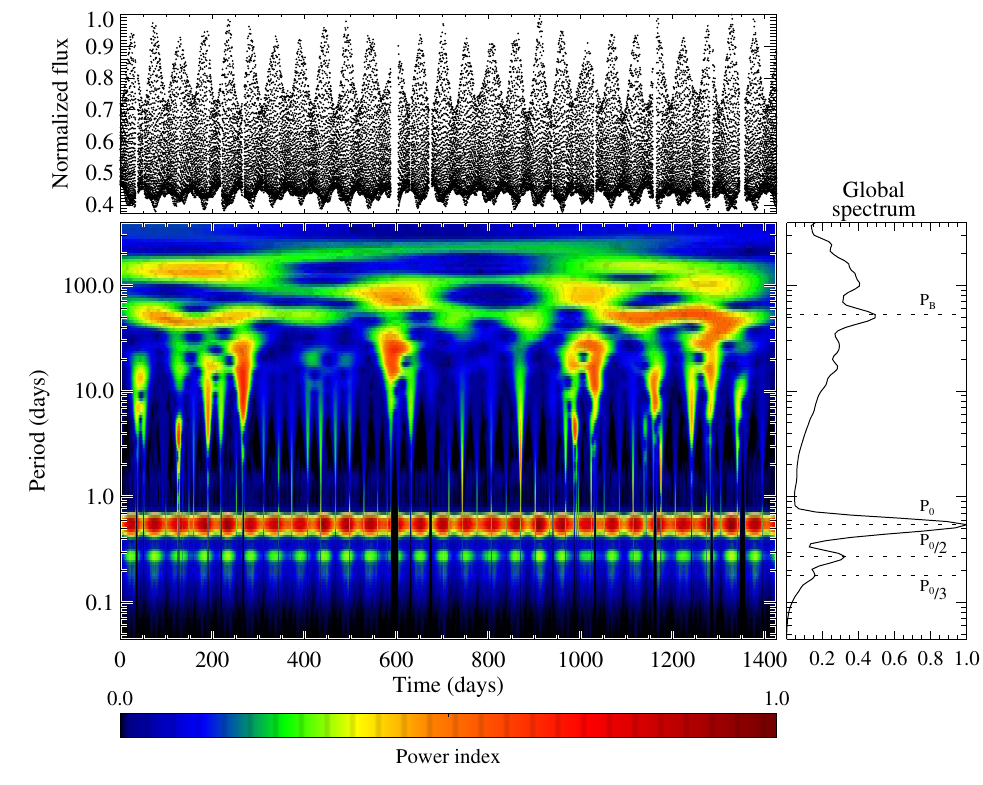}}
\caption{Light curve (at the top) of {\it Kepler} RR Lyrae star KIC 9697825 or V360 Lyr, its local map (in the center) and global wavelet spectrum (to the right). The main periodicities detected are $P_{0} = 0.54$~d, associated to the radial fundamental pulsation period, $\frac{P_{0}}{2} = 0.27$~d and $\frac{P_{0}}{3} = 0.18$~d, the second and third harmonics, and $P_{B} = 52.8$~d, related to the Blazhko period. The 6th-order Morlet wavelet was used.}
\label{wmpV0360}
\end{figure}

To assume this signature as typical of this type of pulsation, we also applied the wavelet analysis to the long-term light curve of the RR Lyrae {\object{KIC 9697825}} (2MASS 19015863+4626457) or V360 Lyr (variable star designation in the GCVS Catalog \citep{2009yCat....102025S}) with a total time span of 1426 days. Figure~\ref{wmpV0360} shows the corresponding WVM and GWS. The contour levels are not plotted here to avoid hiding the pulsation signature in the local map. The beat pattern of the previous RR Lyrae is still evident here; i.e., the dune ranges are circular and regular, comprising tracks associated to the primary period and the harmonics. This beat pattern characterizes the Blazhko cycles (27 full cycles) whose periodic variation could be associated to the 52.8-day periodicity ($P_{B}$ or $f_B = 0.019$~d$^{-1}$) in the GWS. The other periodicities are $P_{0} = 0.54$ d (or $f_0 = 1.852$ d$^{-1}$) corresponding to the radial fundamental pulsation period, and the second and third harmonics $\frac{P_{0}}{2} = 0.27$~d (or $2f_0 = 3.704$~d$^{-1}$) and $\frac{P_{0}}{3} = 0.18$~d (or $3f_0 = 5.556$~d$^{-1}$), respectively. For comparison, we find similar results to those obtained by \citet{2010MNRAS.409.1585B} using Fourier analysis ($f_{0} = 1.79344$~d$^{-1}$ or $P_{0} = 0.55759$~d and $P_{B} = 51.4$~d). The authors also find additional frequencies ($f_{1} = 2.4875$~d$^{-1}$ and $f' = 2.6395$~d$^{-1}$) that we do not obtain using our method, possibly due to their small amplitudes. 
Clearly the wavelet pattern and signatures observed for CoRoT star 105288363, a well defined RRab-type Blazhko RR Lyrae type, as discussed in the previous paragraph, are also observed for KIC 9697825.

\begin{figure}
\resizebox{\hsize}{!}{\includegraphics{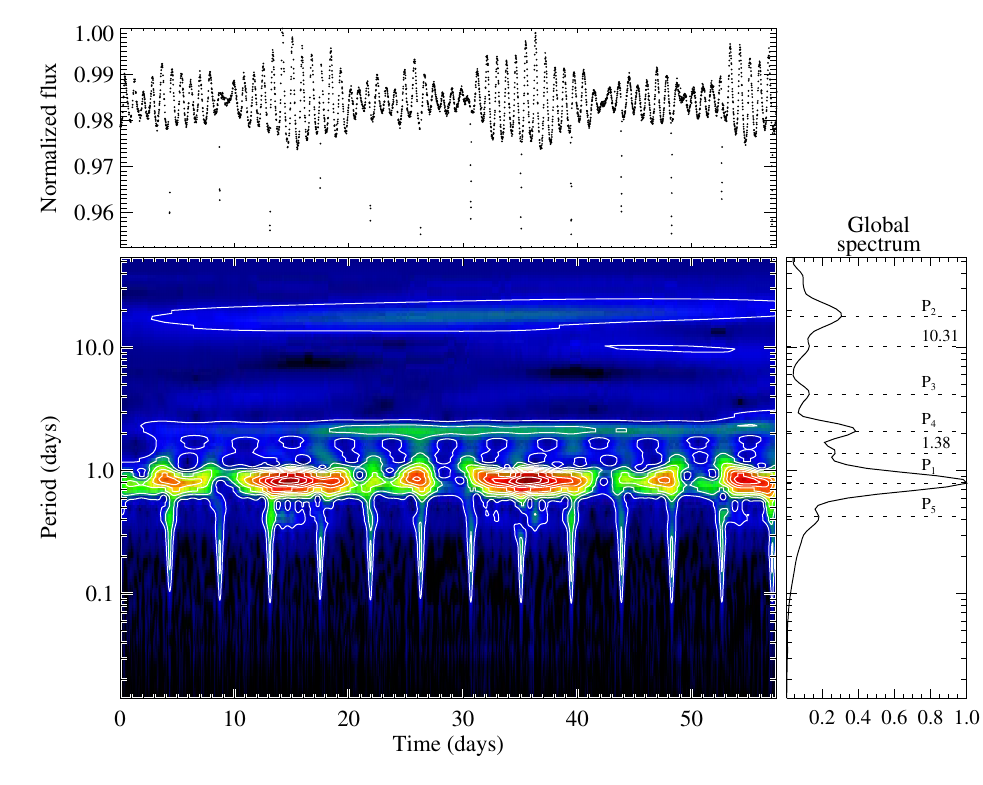}}
\resizebox{\hsize}{!}{\includegraphics{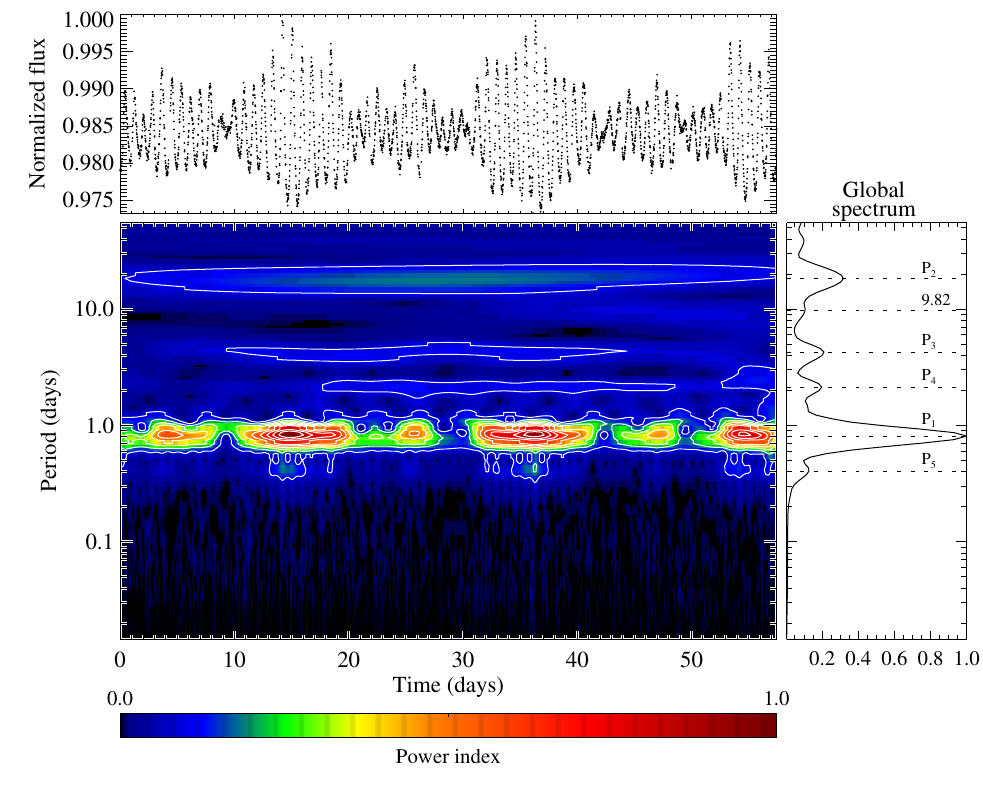}}
\caption{\textit{Top panel}: light curve of CoRoT 102918586 (a $\gamma$ Doradus pulsator) with binary transits (at the top) and its local/global wavelet map (in the center/to the right). \textit{Bottom panel}: light curve of CoRoT 102918586 with eclipses removed (at the top) and its local/global wavelet map (in the center/to the right). The main periodicities detected and illustrated in the GWS are $P_{1} = 0.82$~d, corresponding to the nonradial fundamental pulsation period, $P_{2} = 18.61$~d, related to the beat pattern, $P_{3} = 4.34$~d and $P_{4} = 2.17$~d, also related to the beat pattern (although they are not exactly equal to the harmonics of the $P_{2}$), and $P_{5} = 0.41$~d, an overtone of $P_{1}$. Contour levels are 90\%, 80\%, 70\%,..., 20\% and 10\% of the map maximum. The 6th-order Morlet wavelet was used.}
\label{wmpGDOR}
\end{figure}

The CoRoT star 102918586, observed during the first scientific run anti-center pointing which lasted about 60 d (IRa01), is a 12.4 magnitude eclipsing binary ($ra = 6^{h}48^{m}54.3^{s}$, $dec$ = -0$\degr$52$\arcmin$22.8$\arcsec$, J2000), which is considered to be a $\gamma$ Dor pulsator that shows modulated oscillations and narrow eclipses. For this star, an orbital period of 8.78248 d ($F_{orb} = 0.1139$~d$^{-1}$) is reported by \citet{2010arXiv1004.1525M, 2013A&A...552A..60M}. These authors also present the frequencies obtained from a Fourier analysis and detected a nearly equidistant frequency spacing of about $0.05$ d$^{-1}$. Figure~\ref{wmpGDOR} depicts the wavelet analysis of {\object{CoRoT 102918586}}, considering the binary transits (top panel) and the one with the eclipses removed (bottom panel). There are no significant differences between both maps, with only a few variations in scale intensity caused by the eclipses. However, the pulsation frequency $f_{1} = 1.22$~d$^{-1}$ ($P_{1} = 0.82$~d) is still significant in both WVMs. As shown by \citet{2010arXiv1004.1525M, 2013A&A...552A..60M}, the primary star pulsates with typical $\gamma$ Dor frequencies, a result compatible with our wavelet analysis. The main periodicities illustrated in the GWS of the bottom panel are $P_{1} = 0.82$~d (or $f_{1} = 1.22$~d$^{-1}$), associated to the pulsation period with the highest amplitude, and $P_{2} = 18.61$~d ($f_{2} = 0.05$~d$^{-1}$, corresponding to $\sim{0.5F_{orb}}$) related to the beat pattern. Also, $P_{3} = 4.34$~d ($f_{3} = 0.23$~d$^{-1}$) and $P_{4} = 2.17$~d ($f_{4} = 0.46$~d$^{-1}$) remain after removing the eclipses, leading to the conclusion that they must be related to the beat pattern, although they are not exactly equal to the harmonics of the $P_{2}$. Finally $P_{5} = 0.41$~d ($f_{5} = 2.44$~d$^{-1}$) is one of the harmonics of the pulsation period. All these periodicities are in accordance with those obtained by \citet{2010arXiv1004.1525M}. The 1.38-day periodicity in the GWS of the top panel could be associated to the orbital period because it no longer appears once eclipses are removed, whereas another period of 10.31 days with low power appears in both local maps, assuming that it is also related to the beat pattern of low amplitude. The pulsation signature that we observe here presents a semiregularity of dune ranges in the WVM, putting in doubt that the modulation variations are caused by pulsation or by rotation accompanied by the presence of spots.  Indeed, the observed semiregularity could be the result of the short-term light curve, with a coverage time limited to 57 days, which seems to be very short for identifying evident pulsation signatures by just looking at the local map.

\begin{figure}
\resizebox{\hsize}{!}{\includegraphics{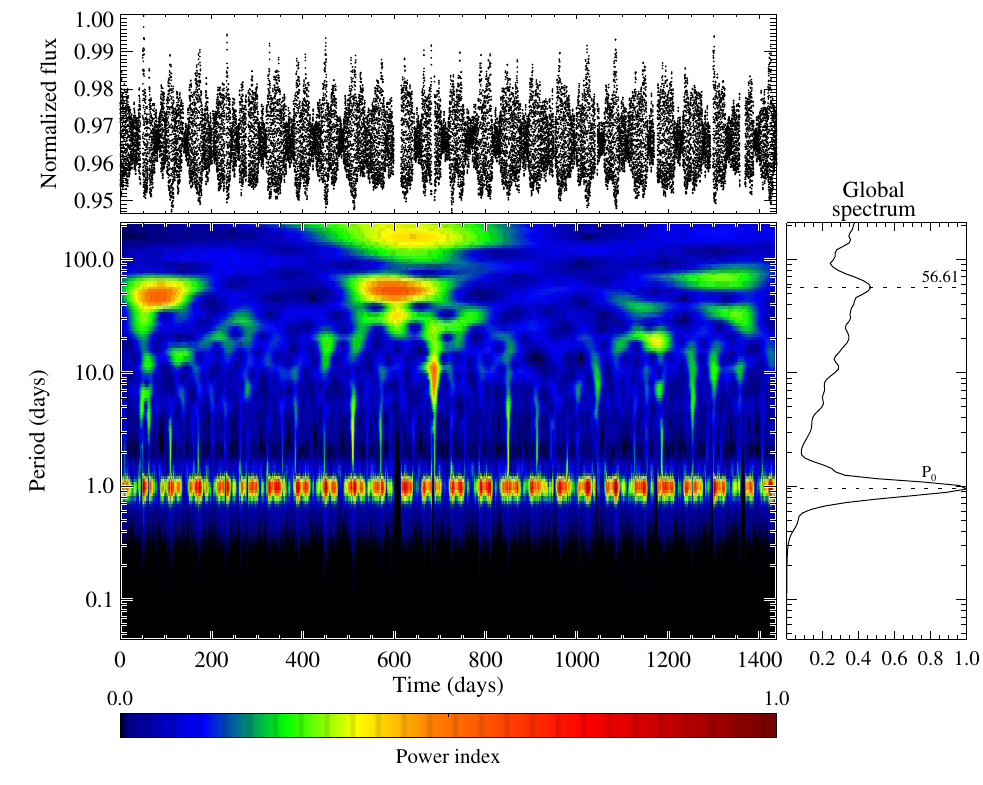}}
\caption{Light curve (at the top) of {\it Kepler} $\gamma$ Dor KIC 3744571, its local map (in the center), and global wavelet spectrum (to the right). The main periodicities detected are $P_{0} = 0.95$~d, associated to the nonradial fundamental pulsation period and $56.61$~d, related to the beat pattern. The 6th-order Morlet wavelet was used.}
\label{wmpKGDOR}
\end{figure}

Finally, we applied the wavelet procedure to the long-term light curve of the {\it Kepler} star {\object{KIC 3744571}} (2MASS 19230559+3848519), classified as a $\gamma$ Dor star by \citet{2013A&A...556A..52T}. From the wavelet analysis, illustrated by the corresponding WVM and GWS given in Fig.~\ref{wmpKGDOR}, we observe one evident track showing a regular continuity of dune ranges associated to pulsation modes. We do not find the same dunes as the previous analyzed RR Lyrae cases because of the difference in the type of pulsations, but the observed regularity points to a clear pulsation pattern. The predominant periods are $P_{0}=0.95$~d, the fundamental pulsation period and 56.61~d, possibly associated to beat pattern, as shown in the GWS. The contour levels are not plotted here to avoid hiding the evidence of $\gamma$ Dor pulsation signature.

\section{Conclusions}
\label{conclusions}

In the present work we have carried out a time--frequency analysis using the wavelet transform of different sorts of stellar light curves obtained in the scope of the CoRoT and {\it Kepler} space missions. The procedure was applied to CoRoT-2 and Kepler-4, two well-known main-sequence stars with planetary transits, to a {\it Kepler} apparently single main-sequence star, KIC 1995351, which is dominated by magnetic activity, to a {\it Kepler} eclipsing binary system KIC 7021177, as well as to four pulsating variable stars, two RR Lyrae, CoRoT 105288363 and KIC 9697825, and two $\gamma$ Dor, CoRoT 102918586 and KIC 3744571. The procedure allowed us to obtain the distribution of the signal’s energy, in time--scale space, from which it was possible to identify the temporal evolution of different phenomena affecting the light curves (such as active regions and possible beats related to pulsations or surface differential rotation).

Wavelet analysis in the solar case gives us a first idea about what to expect if we apply this method to other active stars. As dominant features, reported well by different studies \citep{Howe31032000,2002A&A...394..701K,1990SoPh..130..369D,1999GeoRL..26.3613W}, we identified the 364-day periodicity probably related to an annual solar feature caused by magnetic fluxes generated deep inside the Sun, the solar rotation period, the Rieger-type period, and the 14-day periodicity associated to active regions located opposite each other in solar longitude. The 11-year cycle is also detected when considering the long-term contributions in the local map. 

From the wavelet analysis of the CoRoT-2 light curve, in addition to the orbital period, we identified in particular a main period corresponding to the rotation period and another that is nearly half of the primary period, which is associated to active regions at different longitudes evolving over time. In fact, instead of being considered as an harmonic of the rotation period, this second periodicity could be a presumed effect of active regions moving to the opposite side of the star, most probably due to differential rotation as in solar case \citep{1990SoPh..130..369D}. In the wavelet maps, we also distinguish two semi-regular and continuous dune ranges over the entire time span, which is a strong indicator of the dynamic of starspots yielding to assume these features as the more typical rotation and magnetic activity signature. In addition to these periodicities, a third period is sometimes evident (for example in the case of CoRoT-2) and related to long-term cycles of the stellar activity. However, we can notice that in some cases such a period can be hidden by other contributions in the local map owing to gaps in the light curve (as seen in Fig.~\ref{wmp1177}). In contrast to CoRoT-2, with clear signatures of an active star, the wavelet analysis for Kepler-4 shows no evident signatures of rotation and magnetic activity, reflecting its quiet magnetic activity  behavior, responsible for the observed light curve low amplitude variation. Then, we analyzed KIC 1995351 to compare the identified wavelet signatures with those for CoRoT-2 without transits. In addition to the rotation periodicity, our analysis reveals also the presence of two or more active regions, pointing to a clear dynamic of starspots as in CoRoT-2.

In addition to the confirmation of the orbital period already report in the literature, the wavelet analysis for the {\it Kepler} eclipsing binary KIC 7021177 has also revealed that different periodicity signatures, including rotation, are better defined after removing the transits or eclipses. By comparing the wavelet analysis of the light curves with transits and with removed transits, for stars with planetary and binary transits, it is clear that the presence of periodicity signatures in the light curves are featured more after the transit has been removed, especially when their depths are greater than the amplitude of the rotational modulation.

In the case of the pulsating stars CoRoT 105288363, V360 Lyr, CoRoT 102918586, and KIC 3744571, there are solid similarities between their wavelet maps, both RR Lyrae and $\gamma$ Dor stars clearly showing the pulsation period with its harmonics and a beat pattern illustrated by continuous and regular dune ranges. The pulsation pattern is different between the two types of pulsating stars. The beat pattern for the RR Lyrae stars is represented by more circular and regular dune ranges, whereas the $\gamma$ Dor stars are characterized by more compact but very regular dune ranges (or tracks). We also note that for a short total time span, the $\gamma$ Dor pulsation signature could be confused by rotational modulation. Finally, to establish the observed regularity of dunes as a pulsation pattern in the referred pulsating stars, we extend our wavelet analysis to the following additional pulsating stars: KIC 7257008 RR Lyrae star, KIC 2710594, KIC 3448365, KIC 4547348, KIC 4749989, KIC 10080943, KIC 6462033 $\gamma$ Doradus stars, KIC 9700322 RR-$\delta$ Scuti star, and KIC 3324644 Cepheid star. All these are studies of their pulsating nature reported in the literature by other authors. The resulting wavelet maps confirm that the regularity of dunes in the maps is a major trace of a pulsation pattern.
  
In summary, this study has shown that the wavelet technique offers a detailed interpretation of stellar light curves, giving additional information on different physical phenomena present in the signal. Semi-regular patterns represent changes of active regions due to growth or decay of spots and/or to differential rotation, whereas regular patterns indicate events that are more stable in time, like pulsations. This method has an advantage in relation to the Fourier technique (Lomb-Scargle used), because in addition to identifying transits or eclipses, it is possible to identify the signature of the dynamic of different star characteristics associated to the observed light curves.

\begin{acknowledgements}
The CoRoT space mission was developed and is operated by the French space agency CNES, with the participation of ESA's RSSD and Science Programs, Austria, Belgium, Brazil, Germany, and Spain.
This paper includes data collected by the {\em Kepler} mission. Funding for the {\em Kepler} mission
is provided by the NASA Science Mission directorate.
Some of the data presented in this paper were obtained from the Mikulski Archive for Space Telescopes (MAST). STScI is operated by the Association of Universities for Research in Astronomy, Inc., under NASA contract NAS5-26555. Support for MAST for non-HST data is provided by the NASA Office of Space Science via grant NNX09AF08G and by other grants and contracts.
Research activities at the Federal University of Rio Grande do Norte are supported by continuous grants from the national funding agencies CNPq, and FAPERN and by the INCT-INEspa\c{c}o. J.P.B. acknowledges a CAPES graduate fellowship. S.R. and R.E. acknowledge CNPq undergraduate fellowships. I.C.L. acknowledges a CNPq/PNPD fellowship.  
\end{acknowledgements}

\bibliographystyle{aa}
\bibliography{biblio}

\begin{thebibliography}{43}
\expandafter\ifx\csname natexlab\endcsname\relax\def\natexlab#1{#1}\fi

\bibitem[{{Alonso} {et~al.}(2008){Alonso}, {Auvergne}, {Baglin}, {Ollivier},
  {Moutou}, {Rouan}, {Deeg}, {Aigrain}, {Almenara}, {Barbieri}, {Barge},
  {Benz}, {Bord{\'e}}, {Bouchy}, {de La Reza}, {Deleuil}, {Dvorak}, {Erikson},
  {Fridlund}, {Gillon}, {Gondoin}, {Guillot}, {Hatzes}, {H{\'e}brard},
  {Kabath}, {Jorda}, {Lammer}, {L{\'e}ger}, {Llebaria}, {Loeillet}, {Magain},
  {Mayor}, {Mazeh}, {P{\"a}tzold}, {Pepe}, {Pont}, {Queloz}, {Rauer},
  {Shporer}, {Schneider}, {Stecklum}, {Udry}, \&
  {Wuchterl}}]{2008A&A...482L..21A}
{Alonso}, R., {Auvergne}, M., {Baglin}, A., {et~al.} 2008, \aap, 482, L21

\bibitem[{{Baglin} {et~al.}(2009){Baglin}, {Auvergne}, {Barge}, {Deleuil},
  {Michel}, \& {CoRoT Exoplanet Science Team}}]{2009IAUS..253...71B}
{Baglin}, A., {Auvergne}, M., {Barge}, P., {et~al.} 2009, in IAU Symposium,
  Vol. 253, IAU Symposium, ed. F.~{Pont}, D.~{Sasselov}, \& M.~J. {Holman},
  71--81

\bibitem[{{Baudin} {et~al.}(2006){Baudin}, {Baglin}, {Orcesi}, {Nguyen-Kim},
  {Solano}, {Ochsenbeim}, \& {Corot Scientific
  Committee}}]{2006ESASP1306..145B}
{Baudin}, F., {Baglin}, A., {Orcesi}, J.-L., {et~al.} 2006, in ESA Special
  Publication, Vol. 1306, ESA Special Publication, ed. M.~{Fridlund},
  A.~{Baglin}, J.~{Lochard}, \& L.~{Conroy}, 145

\bibitem[{{Benk{\H o}} {et~al.}(2010){Benk{\H o}}, {Kolenberg}, {Szab{\'o}},
  {Kurtz}, {Bryson}, {Bregman}, {Still}, {Smolec}, {Nuspl}, {Nemec},
  {Moskalik}, {Kopacki}, {Koll{\'a}th}, {Guggenberger}, {di Criscienzo},
  {Christensen-Dalsgaard}, {Kjeldsen}, {Borucki}, {Koch}, {Jenkins}, \& {van
  Cleve}}]{2010MNRAS.409.1585B}
{Benk{\H o}}, J.~M., {Kolenberg}, K., {Szab{\'o}}, R., {et~al.} 2010, \mnras,
  409, 1585

\bibitem[{{Boashash}(1988)}]{1988-Boashash}
{Boashash}, B. 1988, IEEE Transactions on Acoustics, Speech, and Signal
  Processing, 36, 1518

\bibitem[{{Bochner} \& {Chandrasekharan}(1949)}]{Bochner-1949}
{Bochner}, S. \& {Chandrasekharan}, K. 1949, {Fourier Transforms} (Princeton
  University Press)

\bibitem[{Borucki {et~al.}(2010)Borucki, Koch, Basri, Batalha, Brown, Caldwell,
  Caldwell, Christensen-Dalsgaard, Cochran, DeVore, Dunham, Dupree, Gautier,
  Geary, Gilliland, Gould, Howell, Jenkins, Kondo, Latham, Marcy, Meibom,
  Kjeldsen, Lissauer, Monet, Morrison, Sasselov, Tarter, Boss, Brownlee, Owen,
  Buzasi, Charbonneau, Doyle, Fortney, Ford, Holman, Seager, Steffen, Welsh,
  Rowe, Anderson, Buchhave, Ciardi, Walkowicz, Sherry, Horch, Isaacson,
  Everett, Fischer, Torres, Johnson, Endl, MacQueen, Bryson, Dotson, Haas,
  Kolodziejczak, Van~Cleve, Chandrasekaran, Twicken, Quintana, Clarke, Allen,
  Li, Wu, Tenenbaum, Verner, Bruhweiler, Barnes, \& Prsa}]{Borucki19022010}
Borucki, W.~J., Koch, D., Basri, G., {et~al.} 2010, Science, 327, 977

\bibitem[{{Borucki} {et~al.}(2011){Borucki}, {Koch}, {Basri}, {Batalha},
  {Brown}, {Bryson}, {Caldwell}, {Christensen-Dalsgaard}, {Cochran}, {DeVore},
  {Dunham}, {Gautier}, {Geary}, {Gilliland}, {Gould}, {Howell}, {Jenkins},
  {Latham}, {Lissauer}, {Marcy}, {Rowe}, {Sasselov}, {Boss}, {Charbonneau},
  {Ciardi}, {Doyle}, {Dupree}, {Ford}, {Fortney}, {Holman}, {Seager},
  {Steffen}, {Tarter}, {Welsh}, {Allen}, {Buchhave}, {Christiansen}, {Clarke},
  {Das}, {D{\'e}sert}, {Endl}, {Fabrycky}, {Fressin}, {Haas}, {Horch},
  {Howard}, {Isaacson}, {Kjeldsen}, {Kolodziejczak}, {Kulesa}, {Li}, {Lucas},
  {Machalek}, {McCarthy}, {MacQueen}, {Meibom}, {Miquel}, {Prsa}, {Quinn},
  {Quintana}, {Ragozzine}, {Sherry}, {Shporer}, {Tenenbaum}, {Torres},
  {Twicken}, {Van Cleve}, {Walkowicz}, {Witteborn}, \&
  {Still}}]{2011ApJ...736...19B}
{Borucki}, W.~J., {Koch}, D.~G., {Basri}, G., {et~al.} 2011, \apj, 736, 19

\bibitem[{{Bouchy} {et~al.}(2008){Bouchy}, {Queloz}, {Deleuil}, {Loeillet},
  {Hatzes}, {Aigrain}, {Alonso}, {Auvergne}, {Baglin}, {Barge}, {Benz},
  {Bord{\'e}}, {Deeg}, {de La Reza}, {Dvorak}, {Erikson}, {Fridlund},
  {Gondoin}, {Guillot}, {H{\'e}brard}, {Jorda}, {Lammer}, {L{\'e}ger},
  {Llebaria}, {Magain}, {Mayor}, {Moutou}, {Ollivier}, {P{\"a}tzold}, {Pepe},
  {Pont}, {Rauer}, {Rouan}, {Schneider}, {Triaud}, {Udry}, \&
  {Wuchterl}}]{2008A&A...482L..25B}
{Bouchy}, F., {Queloz}, D., {Deleuil}, M., {et~al.} 2008, \aap, 482, L25

\bibitem[{{Burrus} {et~al.}(1998){Burrus}, {Gopinath}, \& {Guo}}]{Burrus-1998}
{Burrus}, C.~S., {Gopinath}, R., \& {Guo}, H. 1998, {Introduction to Wavelets
  and Wavelet Transforms} (Prentice-Hall)

\bibitem[{{Daubechies}(1992)}]{1992tlw..conf.....D}
{Daubechies}, I., ed. 1992, {Ten lectures on wavelets}

\bibitem[{{De Medeiros} {et~al.}(2013){De Medeiros}, {Ferreira Lopes},
  {Le{\~a}o}, {Canto Martins}, {Catelan}, {Baglin}, {Vieira}, {Bravo},
  {Cort{\'e}s}, {de Freitas}, {Janot-Pacheco}, {Maciel}, {Melo}, {Osorio},
  {Porto de Mello}, \& {Valio}}]{2013A&A...555A..63D}
{De Medeiros}, J.~R., {Ferreira Lopes}, C.~E., {Le{\~a}o}, I.~C., {et~al.}
  2013, \aap, 555, A63

\bibitem[{{Dimitrov} {et~al.}(2012){Dimitrov}, {Kyurkchieva}, \&
  {Radeva}}]{2012BlgAJ..18c..81D}
{Dimitrov}, D., {Kyurkchieva}, D., \& {Radeva}, V. 2012, Bulgarian Astronomical
  Journal, 18, 030000

\bibitem[{{Donnelly} \& {Puga}(1990)}]{1990SoPh..130..369D}
{Donnelly}, R.~F. \& {Puga}, L.~C. 1990, \solphys, 130, 369

\bibitem[{{Drake}(2003)}]{2003ApJ...589.1020D}
{Drake}, A.~J. 2003, \apj, 589, 1020

\bibitem[{{Foster}(1996)}]{Foster-1996}
{Foster}, G. 1996, The Astronomical Journal, 112, 1709

\bibitem[{{Fr{\"o}hlich} \& {Lean}(1998)}]{1998GeoRL..25.4377F}
{Fr{\"o}hlich}, C. \& {Lean}, J. 1998, \grl, 25, 4377

\bibitem[{{Fr{\"o}hlich} \& {Lean}(2004)}]{2004A&ARv..12..273F}
{Fr{\"o}hlich}, C. \& {Lean}, J. 2004, \aapr, 12, 273

\bibitem[{{Gabor}(1946)}]{Gabor:JIEEE-93-429}
{Gabor}, D. 1946, J. Inst. Electr. Engineering, 93, 429

\bibitem[{{Garc{\'{\i}}a} {et~al.}(2009){Garc{\'{\i}}a}, {R{\'e}gulo},
  {Samadi}, {Ballot}, {Barban}, {Benomar}, {Chaplin}, {Gaulme}, {Appourchaux},
  {Mathur}, {Mosser}, {Toutain}, {Verner}, {Auvergne}, {Baglin}, {Baudin},
  {Boumier}, {Bruntt}, {Catala}, {Deheuvels}, {Elsworth}, {Jim{\'e}nez-Reyes},
  {Michel}, {P{\'e}rez Hern{\'a}ndez}, {Roxburgh}, \&
  {Salabert}}]{2009A&A...506...41G}
{Garc{\'{\i}}a}, R.~A., {R{\'e}gulo}, C., {Samadi}, R., {et~al.} 2009, \aap,
  506, 41

\bibitem[{{Grossmann} \& {Morlet}(1984)}]{1984-Grossmann-Morlet}
{Grossmann}, A. \& {Morlet}, J. 1984, SIAM Journal Mathematical Analysis, 15,
  723

\bibitem[{{Guggenberger} {et~al.}(2011){Guggenberger}, {Kolenberg},
  {Chapellier}, {Poretti}, {Szab{\'o}}, {Benk{\H o}}, \&
  {Papar{\'o}}}]{2011MNRAS.415.1577G}
{Guggenberger}, E., {Kolenberg}, K., {Chapellier}, E., {et~al.} 2011, \mnras,
  415, 1577

\bibitem[{Howe {et~al.}(2000)Howe, Christensen-Dalsgaard, Hill, Komm, Larsen,
  Schou, Thompson, \& Toomre}]{Howe31032000}
Howe, R., Christensen-Dalsgaard, J., Hill, F., {et~al.} 2000, Science, 287,
  2456

\bibitem[{{Hubbard}(1996)}]{Hubbard-1996}
{Hubbard}, B. 1996, {The world according to wavelets. The Story of a
  Mathematical Technique in the Making } (A. K. Peters Ltd, Wellesley,
  Massachusetts)

\bibitem[{{Jury}(1964)}]{Jury-1964}
{Jury}, E. 1964, {Theory and application of the z-transform method} (Robert E.
  Krieger Publishing)

\bibitem[{{Krivova} \& {Solanki}(2002)}]{2002A&A...394..701K}
{Krivova}, N.~A. \& {Solanki}, S.~K. 2002, \aap, 394, 701

\bibitem[{{Lanza} {et~al.}(2009){Lanza}, {Pagano}, {Leto}, {Messina},
  {Aigrain}, {Alonso}, {Auvergne}, {Baglin}, {Barge}, {Bonomo}, {Boumier},
  {Collier Cameron}, {Comparato}, {Cutispoto}, {de Medeiros}, {Foing},
  {Kaiser}, {Moutou}, {Parihar}, {Silva-Valio}, \&
  {Weiss}}]{2009A&A...493..193L}
{Lanza}, A.~F., {Pagano}, I., {Leto}, G., {et~al.} 2009, \aap, 493, 193

\bibitem[{{Maceroni} {et~al.}(2010){Maceroni}, {Cardini}, {Damiani},
  {Gandolfi}, {Debosscher}, {Hatzes}, {Guenther}, \&
  {Aerts}}]{2010arXiv1004.1525M}
{Maceroni}, C., {Cardini}, D., {Damiani}, C., {et~al.} 2010, ArXiv e-prints

\bibitem[{{Maceroni} {et~al.}(2013){Maceroni}, {Montalb{\'a}n}, {Gandolfi},
  {Pavlovski}, \& {Rainer}}]{2013A&A...552A..60M}
{Maceroni}, C., {Montalb{\'a}n}, J., {Gandolfi}, D., {Pavlovski}, K., \&
  {Rainer}, M. 2013, \aap, 552, A60

\bibitem[{{Mathur} {et~al.}(2014){Mathur}, {Garc{\'{\i}}a}, {Ballot},
  {Ceillier}, {Salabert}, {Metcalfe}, {R{\'e}gulo}, {Jim{\'e}nez}, \&
  {Bloemen}}]{2014A&A...562A.124M}
{Mathur}, S., {Garc{\'{\i}}a}, R.~A., {Ballot}, J., {et~al.} 2014, \aap, 562,
  A124

\bibitem[{{Morettin}(1999)}]{Morettin-1999}
{Morettin}, P.~A. 1999, {Ondas e Ondeletas: da an\'{a}lise de Fourier \`{a}
  an\'{a}lise de ondeletas} (EdUSP)

\bibitem[{{Pr{\v s}a} {et~al.}(2011){Pr{\v s}a}, {Batalha}, {Slawson}, {Doyle},
  {Welsh}, {Orosz}, {Seager}, {Rucker}, {Mjaseth}, {Engle}, {Conroy},
  {Jenkins}, {Caldwell}, {Koch}, \& {Borucki}}]{2011AJ....141...83P}
{Pr{\v s}a}, A., {Batalha}, N., {Slawson}, R.~W., {et~al.} 2011, \aj, 141, 83

\bibitem[{{Reinhold} {et~al.}(2013){Reinhold}, {Reiners}, \&
  {Basri}}]{2013arXiv1308.1508R}
{Reinhold}, T., {Reiners}, A., \& {Basri}, G. 2013, ArXiv e-prints

\bibitem[{{Rieger} {et~al.}(1984){Rieger}, {Kanbach}, {Reppin}, {Share},
  {Forrest}, \& {Chupp}}]{1984Natur.312..623R}
{Rieger}, E., {Kanbach}, G., {Reppin}, C., {et~al.} 1984, \nat, 312, 623

\bibitem[{{Samus} {et~al.}(2009){Samus}, {Durlevich}, \& {et
  al.}}]{2009yCat....102025S}
{Samus}, N.~N., {Durlevich}, O.~V., \& {et al.} 2009, VizieR Online Data
  Catalog, 1, 2025

\bibitem[{{Szab{\'o}}(2014)}]{2014IAUS..301..241S}
{Szab{\'o}}, R. 2014, in IAU Symposium, Vol. 301, IAU Symposium, ed. J.~A.
  {Guzik}, W.~J. {Chaplin}, G.~{Handler}, \& A.~{Pigulski}, 241--248

\bibitem[{{Tenenbaum} {et~al.}(2010){Tenenbaum}, {Bryson}, {Chandrasekaran},
  {Li}, {Quintana}, {Twicken}, \& {Jenkins}}]{2010SPIE.7740E..16T}
{Tenenbaum}, P., {Bryson}, S.~T., {Chandrasekaran}, H., {et~al.} 2010, in
  Society of Photo-Optical Instrumentation Engineers (SPIE) Conference Series,
  Vol. 7740, Society of Photo-Optical Instrumentation Engineers (SPIE)
  Conference Series

\bibitem[{{Tkachenko} {et~al.}(2013){Tkachenko}, {Aerts}, {Yakushechkin},
  {Debosscher}, {Degroote}, {Bloemen}, {P{\'a}pics}, {de Vries}, {Lombaert},
  {Hrudkova}, {Fr{\'e}mat}, {Raskin}, \& {Van Winckel}}]{2013A&A...556A..52T}
{Tkachenko}, A., {Aerts}, C., {Yakushechkin}, A., {et~al.} 2013, \aap, 556, A52

\bibitem[{{Torrence} \& {Compo}(1998)}]{1998BAMS...79...61T}
{Torrence}, C. \& {Compo}, G.~P. 1998, Bulletin of the American Meteorological
  Society, 79, 61

\bibitem[{{Twicken} {et~al.}(2010){Twicken}, {Chandrasekaran}, {Jenkins},
  {Gunter}, {Girouard}, \& {Klaus}}]{2010SPIE.7740E..62T}
{Twicken}, J.~D., {Chandrasekaran}, H., {Jenkins}, J.~M., {et~al.} 2010, in
  Society of Photo-Optical Instrumentation Engineers (SPIE) Conference Series,
  Vol. 7740, Society of Photo-Optical Instrumentation Engineers (SPIE)
  Conference Series

\bibitem[{{Widder}(1945)}]{1945-Widder}
{Widder}, D.~V. 1945, The American Mathematical Monthly, 52, 419

\bibitem[{{Willson}(1997)}]{1997Sci...277.1963W}
{Willson}, R.~C. 1997, Science, 277, 1963

\bibitem[{{Willson} \& {Mordvinov}(1999)}]{1999GeoRL..26.3613W}
{Willson}, R.~C. \& {Mordvinov}, A.~V. 1999, \grl, 26, 3613

\end{thebibliography}

\end{document}